\documentclass[aip,jcp,amsmath,amssymb,floatfix,citeautoscript,reprint]{revtex4-1}

\usepackage{graphicx}
\usepackage{dcolumn}
\usepackage{bm}

\usepackage{braket}
\usepackage[usenames,dvipsnames]{color}

\usepackage{ragged2e}
\usepackage[labelsep=period,format=plain, justification=justified,font=footnotesize]{caption}
\usepackage{subcaption}
\usepackage{times}
\usepackage{txfonts}

\DeclareCaptionJustification{myjust}{\justifying}
\usepackage{xr}
\makeatletter
\newcommand*{\addFileDependency}[1]{
  \typeout{(#1)}
  \@addtofilelist{#1}
  \IfFileExists{#1}{}{\typeout{No file #1.}}
}
\makeatother

\newcommand*{\myexternaldocument}[1]{
    \externaldocument{#1}
    \addFileDependency{#1.tex}
    \addFileDependency{#1.aux}
}

\myexternaldocument{build/si}

\begin{document}

\title{An accurate and efficient Ehrenfest dynamics approach for calculating linear and nonlinear electronic spectra}

\author{Austin O. Atsango}
\affiliation{Department of Chemistry, Stanford University, Stanford, California, 94305, USA}

\author{Andr\'es Montoya-Castillo}
\email{Andres.MontoyaCastillo@colorado.edu}
\affiliation{Department of Chemistry, University of Colorado Boulder, Boulder, Colorado, 80309, USA}

\author{Thomas E. Markland}
\email{tmarkland@stanford.edu}
\affiliation{Department of Chemistry, Stanford University, Stanford, California, 94305, USA}

\date{\today}

\begin{abstract}
Linear and nonlinear electronic spectra provide an important tool to probe the absorption and transfer of electronic energy. Here we introduce a pure state Ehrenfest approach to obtain accurate linear and nonlinear spectra that is applicable to systems with large numbers of excited states and complex chemical environments. We achieve this by representing the initial conditions as sums of pure states and unfolding multi-time correlation functions into the Schr\"odinger picture. By doing this we show that one can obtain significant improvements in accuracy over the previously used projected Ehrenfest approach and that these benefits are particularly pronounced in cases where the initial condition is a coherence between excited states. While such initial conditions do not arise when calculating linear electronic spectra, they play a vital role in capturing multidimensional spectroscopies. We demonstrate the performance of our method by showing that it is able to quantitatively capture the  exact linear, 2DES, and pump-probe spectra for a Frenkel exciton model in slow bath regimes and is even able to reproduce the main spectral features in fast bath regimes.
\end{abstract}

\maketitle
\normalsize

\section{Introduction}

Nonlinear optical spectroscopy is a powerful tool for probing the structure and dynamics of chemical systems~\cite{Ginsberg2009,Cho2008,Khalil2003}, and it typically provides significantly more information than its linear counterparts. For example, 2D electronic spectroscopy (2DES)~\cite{Hybl1998,mukamel1995principles} is frequently used to elucidate inter-chromophoric couplings, energy transfer and relaxation processes, and environmental effects in systems such as photosynthetic molecular aggregates with femtosecond time resolution~\cite{Brixner2005,Engel2007,Dostl2012,Abramavicius2010}. However, interpreting these spectra in terms of the specific molecular structures and motions that give rise to the electronic states and their relaxation pathways often requires the assistance of simulations to disentangle spectral features such as short-time coherent oscillations and long-time population decay pathways. Accurately and efficiently simulating 2DES signals from atomistic simulations and uncovering how they arise from the underlying quantum dynamics of the nuclear and electronic states remains a significant challenge.

Simulating nonlinear optical spectra requires obtaining higher-order response functions and hence the number of methods that have successfully been applied to generate them is significantly more limited than for the less computationally demanding linear spectra. Exact methods such as the Hierarchical Equations of Motion (HEOM)~\cite{Tanimura1989,Ishizaki2007,Kreisbeck2011,Chen2010} and multiconfiguration time-dependent Hartree (MCTDH)~\cite{Meyer2011,Schulze2017}  provide important insights and benchmarks for approximate methods. However, the desire to treat large condensed-phase systems containing many electronic states with a fully atomistic treatment of the nuclear motions and even on-the-fly evaluation of the electronic surfaces has spurred the recent development of approximate dynamics methods. When approximate methods are used, 2DES has been shown to provide a much stricter test of the quantum dynamics method employed than linear electronic spectroscopy. For example, when using Redfield theory\cite{REDFIELD1965,Bloch1957} it has been shown that in parameter regimes where the linear electronic spectrum can be quantitatively captured, the 2DES spectrum significantly deviates from the exact results~\cite{Fetherolf2017}, although this can be somewhat alleviated by freezing the low-frequency bath modes.\cite{MontoyaCastillo2015,Fetherolf2017}

Trajectory-based methods provide a particularly appealing approach to simulate 2DES since they are typically compatible with a fully atomistic treatment of the nuclear motions and even on-the-fly evaluation of the electronic surfaces on which the nuclei evolve. Many recent applications of trajectory-based methods to 2DES have been based on semiclassical treatments of the Meyer-Miller-Stock-Thoss~\cite{Meyera1979,Stock1997} mapping of the electronic degrees of freedom. In particular, the partially linearized density matrix (PLDM) approach~\cite{Huo2011} and its more recent spin-PLDM variant~\cite{Mannouch-2020,Mannouch2020}  have been shown to produce accurate 2DES spectra in parameter regimes where perturbative methods break down~\cite{Provazza2018,Mannouch2022}. Further work has shown how mapping-based semiclassical methods can be used to simulate 2DES beyond the perturbative treatment of the field-matter interaction~\cite{Gao2020}. The optimized mean trajectory approach has also been recently coupled with trajectories based on the Meyer-Miller Hamiltonian to compute two-dimensional vibrational-electronic spectra~\cite{Polley2021,Loring2022}. Other trajectory-based approaches include the numerical integration of the Schrodinger equation (NISE)~\cite{Jansen2006,Liang2012,Torii2006} and the stochastic Liouville Equation methods~\cite{Jansen2004,Jansen2005}, which involve an explicit treatment of the bath degrees of freedom but do not account for the back-reaction of the electronic degrees of freedom on the bath, leading to inaccuracies in linear and nonlinear spectra. To correct this, NISE has been combined with the fewest switches surface hopping approach, allowing it to incorporate the quantum back-reaction on the bath and thus improve its accuracy.\cite{Tempelaar2013}

Here, we present an accurate method to simulate 2DES using Ehrenfest dynamics and contrast it with a previously reported method detailed in Ref.~\onlinecite{vanderVegte2013}. By applying our method, we show that significantly more accurate 2DES spectra can be obtained from Ehrenfest dynamics if one uses an appropriate initialization of coherence and mixed states. In particular, we express all initial conditions as sums of pure states, which have important implications for the accuracy of the resulting dynamics and allow the simulation to be done unambiguously in either the wavefunction or density matrix formulations of Ehrenfest theory. We show that when formulated in this way, Ehrenfest theory can closely reproduce the HEOM result, and that this result is not achieved when initialization is not done from pure states.

\section{Pure State and Projected Ehrenfest Methods} \label{sec:method}

It has previously been shown that in the density matrix formulation of the Ehrenfest method~\cite{Kapral1999}, a carefully constructed initialization is required to obtain unambiguous results because density matrices initialized from non-pure states do not yield unique dynamics~\cite{MontoyaCastillo2016}. Conversely, the wavefunction formulation of Ehrenfest does not suffer from this problem because all initial conditions are automatically pure states. Our approach to the density matrix formulation of Ehrenfest avoids this previously reported ambiguity by expanding any given initial density matrix $\rho_0$ as a sum of pure states,
\begin{equation} \label{sum_ps}
    \rho_0 = \sum_{\rm{i}} a_{\rm{i}} \ket{\psi_{\rm{i}}}\bra{\psi_{\rm{i}}}.
\end{equation}
For an arbitrary initial density matrix $\rho_0 = \ket{a}\bra{b}$, the pure state decomposition can be accomplished as follows:
\begin{equation}
\ket{a}\bra{b} = \frac{1}{2} \left( \rho^+_{\rm{ab}} + i\rho^-_{\rm{ab}} - \left[1+i\right] \left( \rho_{\rm{aa}} + \rho_{\rm{bb}} \right) \right)
\end{equation}
where
\begin{equation}
    \begin{split}
        \rho_{\rm{aa}} &= \ket{a}\bra{a} \\
        \rho_{\rm{bb}} &= \ket{b}\bra{b} \\
        \rho^+_{\rm{ab}} &= \frac{1}{2}\left( \ket{a} + \ket{b} \right) \left( \bra{a} + \bra{b} \right) \\
        \rho^-_{\rm{ab}} &= \frac{1}{2}\left( \ket{a} + i \ket{b} \right) \left( \bra{a} - i \bra{b} \right)
    \end{split}
\end{equation}
are pure states. Here we will refer to this approach as pure state Ehrenfest and contrast it with the unmodified alternative, which we refer to as projected Ehrefest, where all initial conditions $\ket{a}\bra{b}$ are propagated directly without having been decomposed into pure states.

\section{Simulation Details: Frenkel exciton model} \label{sec:model}

We consider a Frenkel exciton model of coupled chromophores where the full matter Hamiltonian is
\begin{equation} \label{full_ham}
    \hat{H}_{\rm{mat}} = \hat{H}_{\rm{s}} + \hat{H}_{\rm{b}} + \hat{H}_{\rm{sb}}.
\end{equation}
Here, $\hat{H}_{\rm{s}}$ is the system Hamiltonian, $\hat{H}_{\rm{b}}$ is the bath Hamiltonian, and $\hat{H}_{\rm{sb}}$ is the system-bath coupling. The system Hamiltonian, $\hat{H}_{\rm{s}}$, consists of individual chromophores, each with a site energy $\epsilon_{\rm{m}}$, that couple electronically via the transfer integrals $J_{\rm{mn}}$:
\begin{equation}
    \hat{H}_{\rm{s}} = \sum_{\rm{m}} \epsilon_{\rm{m}} \ket{m}\bra{m} + \sum_{\rm{m \neq n}} J_{\rm{mn}} \ket{m}\bra{n}.
\end{equation}
The bath Hamiltonian, $\hat{H}_{\rm{b}}$, consists of independent sets of phonon modes coupled to each chromophore. It is expressed in terms of the phonon mode frequencies $\omega$ and their mass-weighted momenta and coordinates $\hat{P}$ and $\hat{Q}$ as:
\begin{equation}
    \hat{H}_b = \sum_{\rm{m}} \sum_{\rm{j=1}}^{\rm{N^m_b}}\left( \frac{1}{2} \hat{P}^2_{\rm{mj}} + \frac{1}{2} \omega^2_{\rm{mj}}\hat{Q}^2_{\rm{mj}}\right).
\end{equation}
Each chromophore site $\ket{m}\bra{m}$ is linearly coupled to its local bath of phonon modes, such that $\hat{H}_{\rm{sb}}$ takes the form
\begin{equation}
    \hat{H}_{sb} = \sum_{\rm{m}} \sum_{\rm{j=1}}^{\rm{N^m_b}}c_{\rm{mj}}\hat{Q}_{\rm{mj}}\ket{m}\bra{m},
\end{equation}
where $c_{\rm{mj}}$ is the coupling strength of the $j^{th}$ phonon mode attached to chromophore $m$. The characteristics of the baths and their effect on the chromophores are specified via the spectral density,
\begin{equation}
    J_{\rm{m}}(\omega) = \frac{\pi}{2} \sum_{\rm{j}} \frac{c^2_{\rm{mj}}}{\omega_{\rm{j}}} \delta (\omega - \omega_{\rm{j}}).
\end{equation}
All chromophore sites are assumed to have identical spectral densities ($J_{\rm{m}}(\omega) = J (\omega)$). Here, we use an Ohmic spectral density with a Lorentzian cutoff (Debye),
\begin{equation}
    J(\omega) = \frac{2 \lambda \omega_{\rm{c}} \omega}{\omega_{\rm{c}}^2 + \omega^2},
\end{equation}
where $\omega_{\rm{c}}$ is the bath cut-off frequency and~$\lambda = (\hbar \pi)^{-1} \int_0^\infty d\omega J(\omega)/\omega$ is the reorganization energy.

To obtain linear and non-linear spectra, we treat the light-matter interaction perturbatively~\cite{mukamel1995principles}, i.e.,
\begin{equation}
    \hat{H}_{\rm{spec}} = \hat{H}_{\rm{mat}} - \bm{\mu} \cdot \bm{E(t)},
\end{equation}
where $\bm{E(t)}$ is the classical electric field and $\bm{\mu}$ is the total dipole operator,
\begin{equation}
    \bm{\mu} = \sum_{\rm{m}} \mu_{\rm{m}} (\ket{m}\bra{0} + \ket{0}\bra{m}).
\end{equation}

We consider a system of two coupled chromophores with $\epsilon_1 = 50$~cm$^{-1}$, $\epsilon_2 = -50$~cm$^{-1}$, and $J_{12} =$~ 100~cm$^{-1}$, consistent with that used in previous studies~\cite{Ishizaki2009,Fetherolf2017}. In this system, there are four states, all of which are accessible: the ground state, two singly excited states, and one doubly excited state. We report results for two sets of bath parameters: a slow bath with a relaxation time $\omega_c^{-1} = 300$ fs and a fast bath with a relaxation time $\omega_c^{-1}$ = 17.7 fs, both at $k_BT = 208$~cm$^{-1}$. Each bath was discretized using 300 phonon modes via the method employed in Ref.~\onlinecite{Berkelbach2012} that is designed to yield the exact reorganization energy regardless of the number of modes used. The reorganization energy $\lambda$ was set to 50~cm$^{-1}$. All spectra were calculated with a transition dipole matrix where $\mu_1/ \mu_2 = -5$. The slow bath parameter regime was previously investigated via HEOM and Redfield theory and its frozen mode variant in Ref.~\onlinecite{Fetherolf2017}. 

Our HEOM and projected Ehrenfest calculations were conducted using the python package pyrho~\cite{https://doi.org/10.5281/zenodo.4015527}. For the HEOM calculations, converged results for the regimes studied here were obtained by employing the high-temperature approximation, i.e. using zero Matsubara frequencies ($K=0$), and truncating the auxiliary density matrices at $L=15$. Both the HEOM and projected Ehrenfest equations of motion were propagated using a fourth-order Runge-Kutta integrator, while the pure state Ehrenfest method was propagated via the split Liouvillian method, allowing for the use of larger time steps. In the slow bath parameter regime, we used time steps of 10~fs for HEOM and 2~fs for both versions of Ehrenfest, while in the fast bath parameter regime, we used a time step of 2.5~fs for HEOM and 10~fs for pure state Ehrenfest. The linear spectra were generated using 20000 trajectories of length 400 fs, while the nonlinear spectra used 20000 trajectories of length 300 fs for the slow bath regime and 200 fs for the fast bath regime.

\section{Results}
\label{sec:results}

Here we compare the Ehrenfest results for both the linear and nonlinear optical spectra of the Frenkel exciton model outlined in Sec.~\ref{sec:model} to spectra obtained using the numerically exact HEOM approach. We demonstrate that, while in linear absorption spectra the differences between pure state and projected Ehrenfest are subtle, they become much more pronounced for nonlinear spectra.

\subsection{Linear Optical Spectroscopy} \label{sec:linear}

\begin{figure}[h!]
    \centering
    \vspace{-4mm}
    \includegraphics[width=0.3\textwidth]{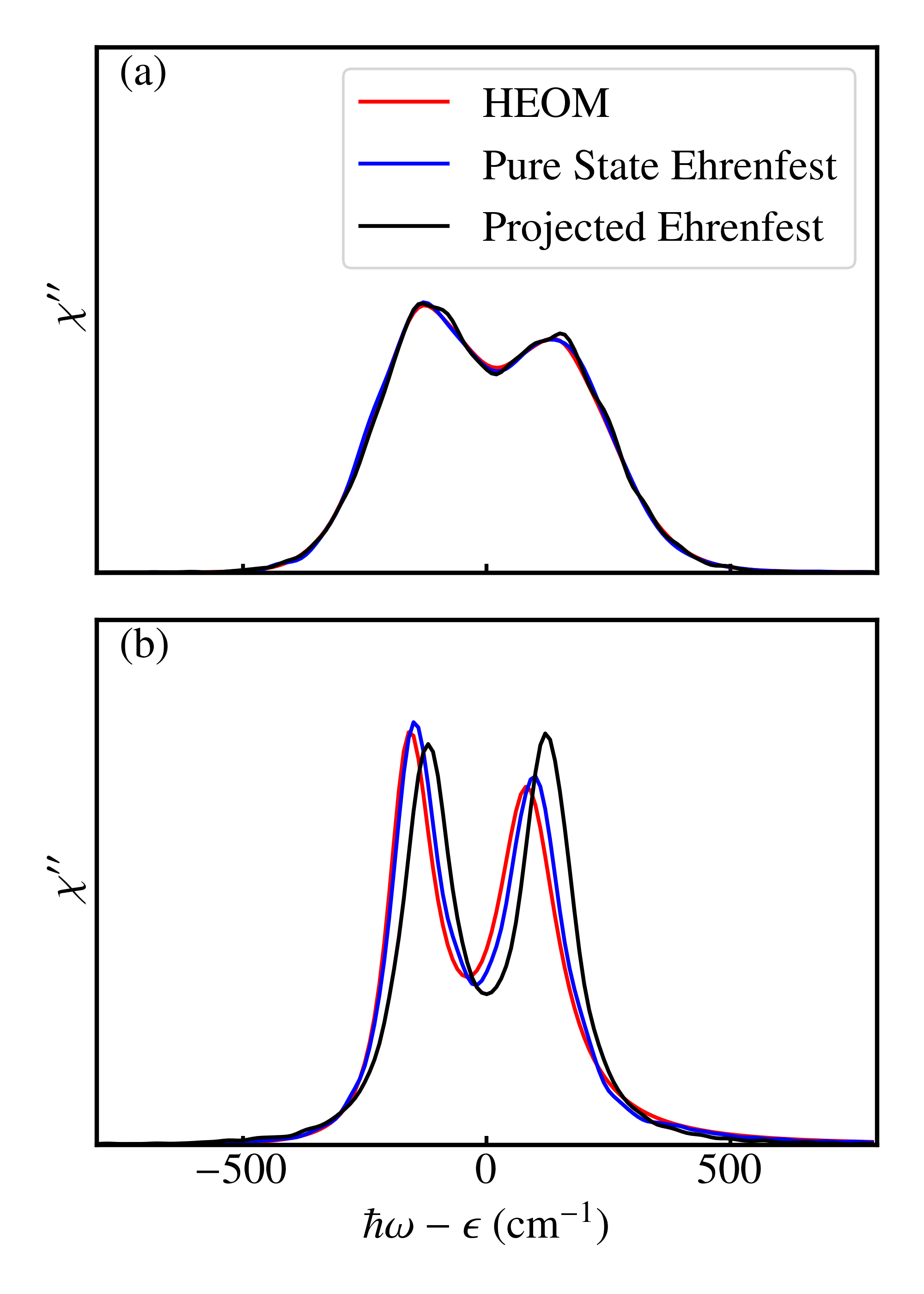}
    \vspace{-4mm}
    \caption{Linear absorption spectra for the slow bath (a) and fast bath (b) parameter regimes computed via HEOM, pure state Ehrenfest, and projected Ehrenfest.}
    \vspace{-4mm}
    \label{fig:abs_spectra}
\end{figure}

\begin{figure}[h!]
    \centering
    \vspace{-6mm}
    \includegraphics[width=0.3\textwidth]{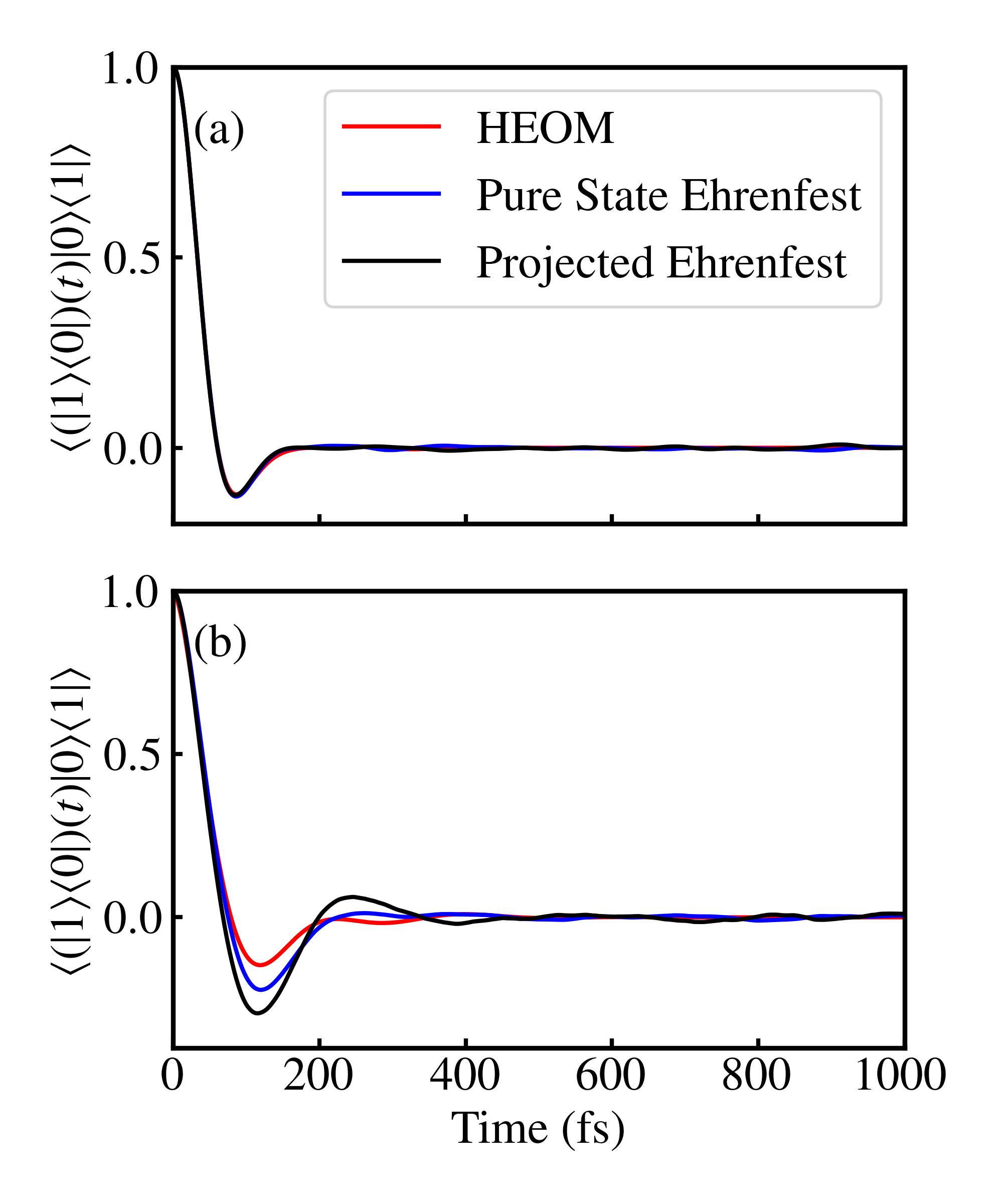}
    \vspace{-4mm}
    \caption{Coherence dynamics for the slow bath (a) and fast bath (b) parameter regimes computed via HEOM, pure state Ehrenfest, and projected Ehrenfest.}
    \vspace{-4mm}
    \label{fig:coh_dynamics}
\end{figure}

Linear absorption spectra can be obtained from the Fourier transform of the first-order response function, $\chi(t)$~\cite{mukamel1995principles},
\begin{equation}
\begin{split}
    \chi(t) = \mathrm{Tr}[\hat{\mu}e^{-i\hat{H}t}\hat{\mu}(0)\rho(0)e^{i\hat{H}t}] \\
    \sigma(\omega) = \int_0^\infty dt e^{i\omega t} \chi(t).
\end{split}
\end{equation}
Applying the dipole operator at $t=0$ results in an optical coherence between the ground state and singly excited states, $\tilde{\rho}(0) = \hat{\mu}(0)\rho(0)$. This initial condition can be represented either as an unmodified projected state (i.e., $\ket{e}\bra{g}$) or as a sum of pure states. Absorption spectra computed via pure state Ehrenfest, projected Ehrenfest, and the numerically exact HEOM for both the fast and slow bath parameter regimes are shown in Fig.~\ref{fig:abs_spectra}. In the slow bath regime (Fig.~\ref{fig:abs_spectra} (a)), where Ehrenfest theory is expected to be accurate, both projected Ehrenfest and pure state Ehrenfest quantitatively reproduce the HEOM result. In the fast bath parameter regime (Fig.~\ref{fig:abs_spectra} (b)), despite qualitative agreement with the HEOM result, there are small inaccuracies for both Ehrenfest methods, with the pure state Ehrenfest method giving a more accurate result than the projected Ehrenfest method. These results can be explained by examining the dynamics arising from the coherence initial condition, $\rho_S(0) = \ket{0}\bra{1}$ shown in Figs.~\ref{fig:coh_dynamics} (a) and (b). $\ket{0}\bra{1}$ is one of the coherence initial conditions (the others being $\rho_S(0) = \ket{0}\bra{2}$ and their complex conjugates) that is propagated to give rise to the first-order response function, $\chi(t)$. In the slow bath regime, there is near quantitative agreement between both versions of Ehrenfest and the HEOM result, while in the fast bath regime, there are more pronounced differences, with the pure state Ehrenfest dynamics matching the HEOM result more closely and the projected Ehrenfest result being underdamped. These differences mirror and point to the source of inaccuracies in the absorption spectra for the fast bath parameter regime. The overall similarity in coherence dynamics across different methods (Fig.~\ref{fig:coh_dynamics}) also reveals why the linear spectra appear to be relatively insensitive to the initialization scheme used for Ehrenfest dynamics: the dynamics starting from a density matrix corresponding to \textit{a coherence with the ground state} are not as sensitive to the initialization scheme. The relative insensitivity of the linear absorption spectra to methods that inaccurately treat coherence dynamics has previously been observed for Redfield theory, which was shown to yield accurate linear absorption spectra despite getting incorrect population dynamics~\cite{Fetherolf2017}, albeit with a different dynamical method rooted in perturbation theory.

\subsection{Nonlinear Optical Spectroscopy} \label{sec:nonlinear}

Two-dimensional electronic spectra offer a much stricter test than absorption spectroscopy of a method's accuracy because they require one to correctly capture both the population and coherence dynamics for a wider range of initial conditions.

\begin{figure}[h]
    \centering
    \includegraphics[width=0.5\textwidth]{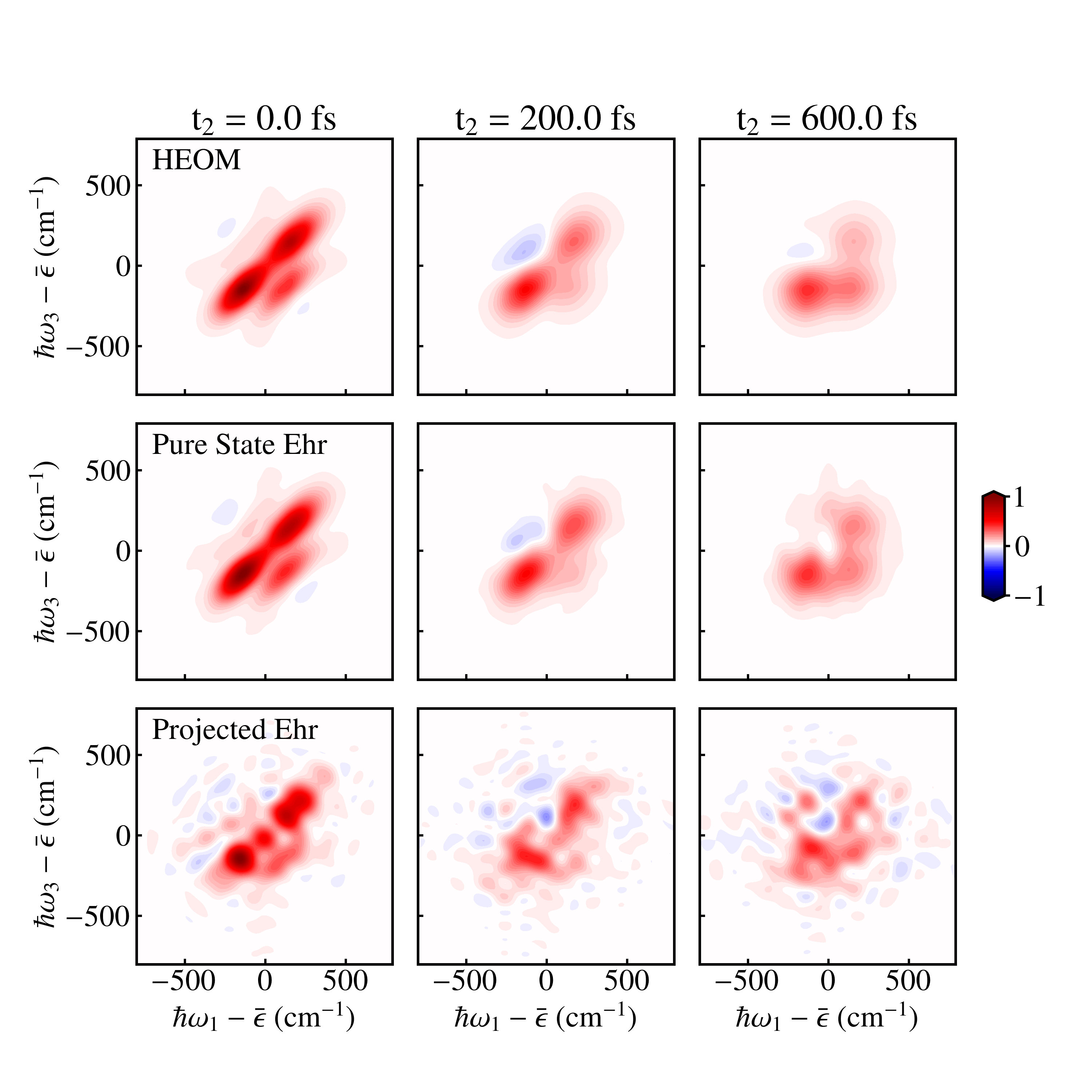}
    \caption{Exact 2DES spectra for the slow bath parameter regime computed using HEOM (top row). 2DES spectra computed using the pure state Ehrenfest and projected Ehrenfest methods are shown in the middle and bottom rows respectively. All spectra are normalized such that the maximum amplitude equals 1.}
    \label{fig:2D_spec_slow}
\end{figure}

2DES spectra can be obtained from a Fourier transform over the rephasing ($R_{rp}$) and non-rephasing ($R_{nr}$) terms of third-order response function under the rotating wave approximation~\cite{mukamel1995principles},
\begin{equation} \label{eq:3o_response}
    \begin{split}
    S(\omega_3, t_2, \omega_1) &= \mathrm{Re} \int_0^\infty dt_1 \int_0^\infty dt_3 [ e^{i(\omega_1 t_1 + \omega_3)} R_{\rm{nr}}(t_3,t_2,t_1) \\
    &+ e^{i(-\omega_1 t_1 + \omega_3)}R_{\rm{rp}}(t_3,t_2,t_1)],
    \end{split}
\end{equation}
where
\begin{equation} \label{eq:rp_nr}
    \begin{split}
    R_{\rm{rp}} &= \Phi_1(t_3,t_2,t_1) + \Phi_2(t_3,t_2,t_1) - \Phi_3(t_3,t_2,t_1), \\
    R_{\rm{nr}} &= \Phi_4(t_3,t_2,t_1) + \Phi_5(t_3,t_2,t_1) - \Phi_6(t_3,t_2,t_1)
    \end{split}
\end{equation}
include the ground state bleaching, stimulated emission, and excited state absorption contributions, and
\begin{equation} \label{eq:response_cont}
    \begin{split}
        \Phi_1(t_3,t_2,t_1) &= \langle \mu_- \mathcal{G}(t_3)[\mathcal{G}(t_2)[ \mu_+ \mathcal{G}(t_1)( \rho_0 \mu_-)] \mu_+] \rangle \\
        \Phi_2(t_3,t_2,t_1) &= \langle \mu_- \mathcal{G}(t_3)[ \mu_+ \mathcal{G}(t_2)[ \mathcal{G}(t_1)( \rho_0 \mu_-) \mu_+]] \rangle \\
        \Phi_3(t_3,t_2,t_1) &= \langle \mu_- \mathcal{G}(t_3)[ \mu_+ \mathcal{G}(t_2)[ \mu_+ \mathcal{G}(t_1)( \rho_0 \mu_-)]] \rangle \\
        \Phi_4(t_3,t_2,t_1) &= \langle \mu_- \mathcal{G}(t_3)[\mathcal{G}(t_2)[ \mathcal{G}(t_1)( \mu_+ \rho_0) \mu_-] \mu_+] \rangle \\
        \Phi_5(t_3,t_2,t_1) &= \langle \mu_- \mathcal{G}(t_3)[\mu_+ \mathcal{G}(t_2)[ \mu_- \mathcal{G}(t_1)( \mu_+ \rho_0 )]] \rangle \\
        \Phi_6(t_3,t_2,t_1) &= \langle \mu_- \mathcal{G}(t_3)[ \mu_+ \mathcal{G}(t_2)[ \mathcal{G}(t_1)( \mu_+ \rho_0 )] \mu_-] \rangle.
    \end{split}
\end{equation}
Here, $\mathcal{G}(t) \rho = e^{-iH_{\rm{mat}}t/\hbar} \rho e^{iH_{\rm{mat}}t/\hbar}$ is the Liouville-space propagator and
\begin{equation}
    \begin{split}
        \mu_- &= \sum_{\rm{m}} \mu_{\rm{m}} \ket{0}\bra{m} \\
        \mu_+ &= \sum_{\rm{m}} \mu_{\rm{m}} \ket{m}\bra{0}.
    \end{split}
\end{equation}
A procedure for obtaining the third-order response function from Ehrenfest dynamics was previously outlined in Ref.~\onlinecite{vanderVegte2013}, which employs ensemble averages over the ground state wavefunction, $\ket{0}$. In this scheme, the dipole operator is applied via a product with individual dipole matrix elements, time propagation is done only over select population states, and the effect of coherences is modeled via an average over two wavefunctions formed from the ket and bra. However, the use of specific dipole matrix elements as well as time-propagation over select population states make it difficult to generalize that method to arbitrary systems with multiple quantum states. Here, we present a generalizable method that exploits the linearity of density matrices to compute third-order response functions as follows:
\begin{enumerate}
    \item Apply the first dipole operator to $\rho_0$ at $t=0$ and split the resultant density matrix into 4 pure states as described in Section~\ref{sec:method}. Via the Condon approximation, the bath degrees of freedom are untouched by the dipole operator and are initially sampled from the Wigner transform of the Boltzmann distribution on the ground state. Each resultant pure state inherits its own copy of the bath. Propagate states and their corresponding baths independently through $t_1$.
    \item Apply the second dipole operator at $t=t_1$, this time to all 4 propagated states. This operation usually results in a new set of non-pure states, which are subsequently split into 4 of their respective pure states, bringing the total number of branches to 16. Each branch adopts a copy of bath coordinates and momenta from its parent branch, ensuring continuity from $t=0$. Propagate the pure states and baths independently through the waiting time, $t_2$.
    \item Apply the third dipole operator at $t=t_1+t_2$ to the 16 propagated states. This splits them again into a total of 64 branches, each a pure state. Propagate all branches through $t_3$.
    \item Consolidate all 64 branches at $t=t_1+t_2+t_3$ and apply the final dipole operator to obtain the third-order response function.
\end{enumerate}
Multiple instances of this procedure are run in order to average over an ensemble of initial bath states. From the above, it is clear that the pure state Ehrenfest method requires more sampling than projected Ehrenfest, which does not involve any splitting of initial conditions into pure states. As we will see below, this splitting into pure states is required to obtain accurate 2DES spectra. The procedure can also be trivially parallelized since all trajectories in the scheme run independently. Additionally, as demonstrated in Section~\ref{sec:method}, we decompose each initial density matrix $\ket{a}\bra{b}$ as a sum of only 4 pure states in the same Liouville space. 

An alternative method for obtaining accurate Ehrenfest dynamics from non-pure state density matrices is outlined in our previous work in the Appendices of Ref.~\onlinecite{Pfalzgraff2019}, and it involves stochastically sampling an auxiliary wavefunction. For example, the initial condition $\ket{0}\bra{1}$ would be obtained from
\begin{equation}
    \ket{\psi} = \frac{1}{\sqrt{2}} \left( \ket{0} + e^{i \phi} \ket{1} \right),
\end{equation}
where $\phi$ is a parameter that is sampled over the range $0 - 2\pi$. In this stochastic scheme, sampling is required both over values of $\phi$ and initial positions and momenta of the bath. In the context of a third-order response function, $\phi$ would require fresh sampling every time a measurement is made (i.e., at times $t=0$, $t=t_1$, and $t=t_1+t_2$) because the auxiliary wavefunction would otherwise return incorrect dynamics. In contrast, the pure state Ehrenfest method detailed above only requires the usual sampling over initial bath conditions, which can be done once at $t=0$ since the baths are continuous through $t=t_1+t_2+t_3$.

\begin{figure}[h!]
    \centering
    \vspace{-8mm}
    \includegraphics[width=0.5\textwidth]{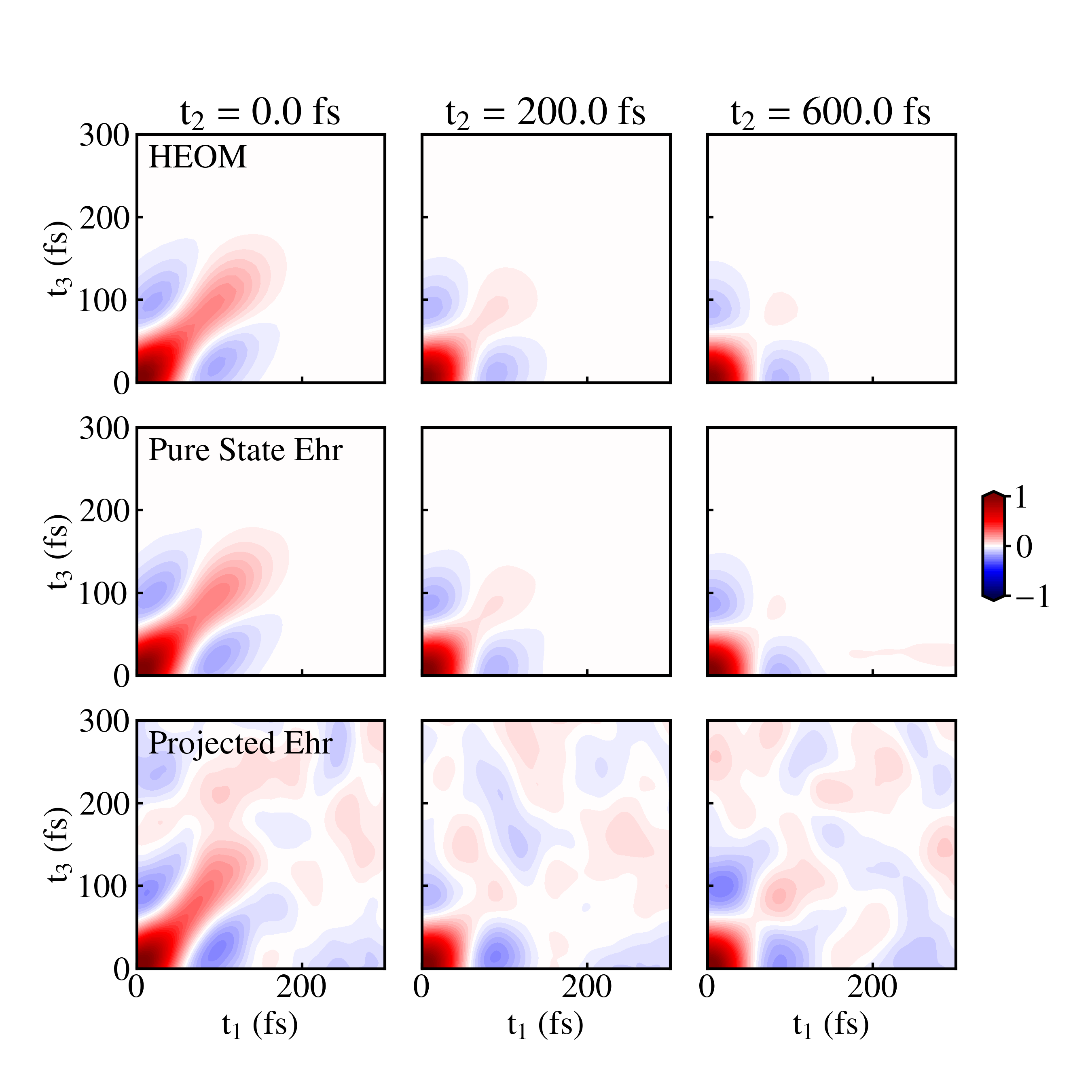}
    \vspace{-6mm}
    \caption{Real parts of $\Phi_2$ for the slow bath parameter regime as computed via HEOM, pure state Ehrenfest, and projected Ehrenfest.}
    \label{fig:phi_2_real}
    \vspace{-4mm}
\end{figure}

\begin{figure}[h!]
    \centering
    \includegraphics[width=0.5\textwidth]{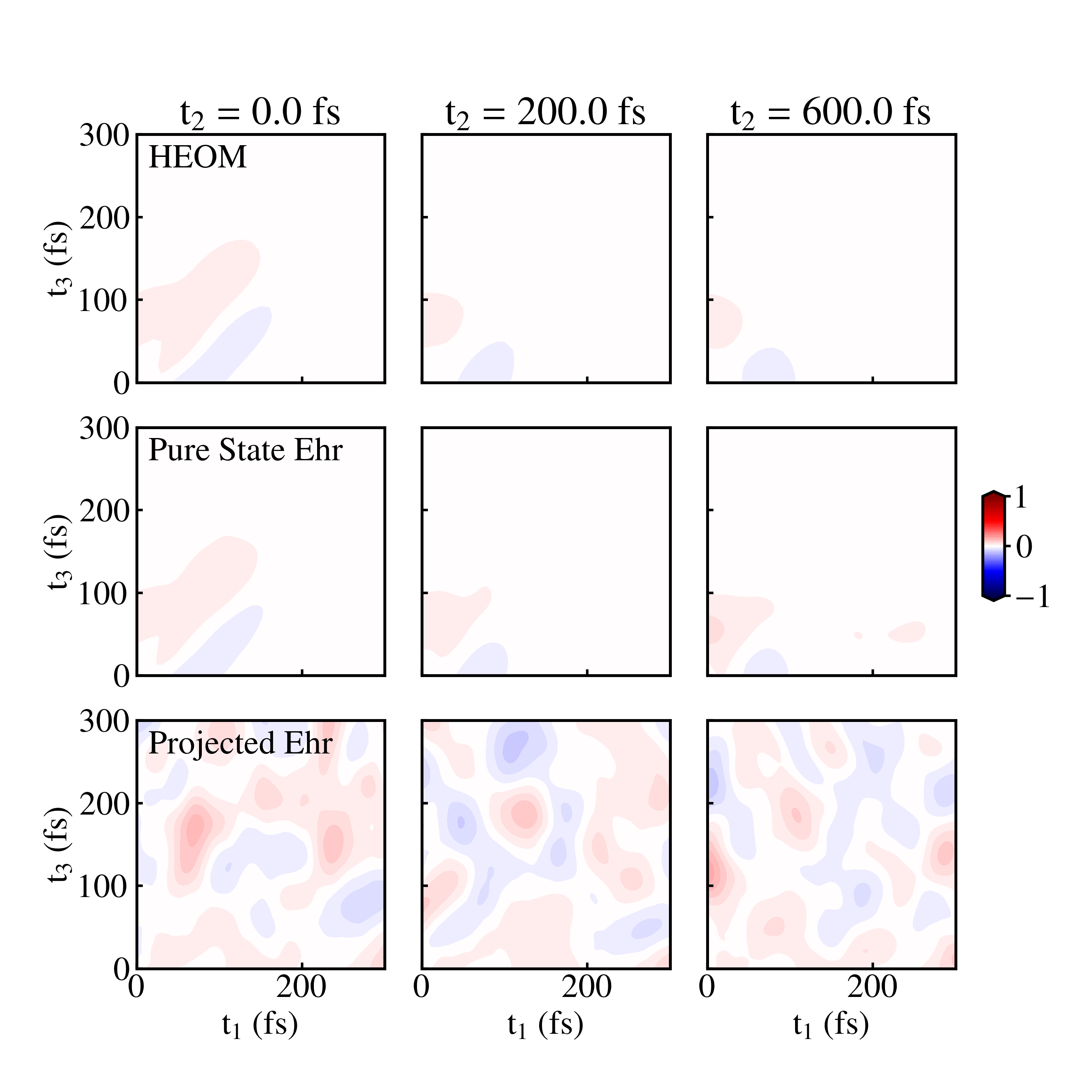}
    \vspace{-4mm}
    \caption{Imaginary parts of $\Phi_2$ for the slow bath parameter regime as computed via HEOM, pure state Ehrenfest, and projected Ehrenfest.}
    \label{fig:phi_2_imag}
    \vspace{-4mm}
\end{figure}

Fig.~\ref{fig:2D_spec_slow} shows the 2DES spectra obtained from both the pure state and projected forms of Ehrenfest and how they compare to the HEOM spectra. The response functions used to obtain the Ehrenfest spectra were truncated (windowed using a step function) as outlined in the SI Sec.~\ref{sec:third_order_truncation}. From this, we observe that pure state Ehrenfest recovers almost quantitative agreement with the HEOM result, with just a slight deterioration as the waiting time $t_2$ increases. In contrast, projected Ehrenfest yields much worse agreement with the HEOM spectrum. While most of the spectral features from the HEOM spectrum are present at $t_2=0$, the projected Ehrenfest spectrum becomes progressively worse as $t_2$ increases and bears little resemblance to the HEOM result by the time $t_2 = 600$ fs.

\begin{figure}[h!]
    \centering
    \includegraphics[width=0.5\textwidth]{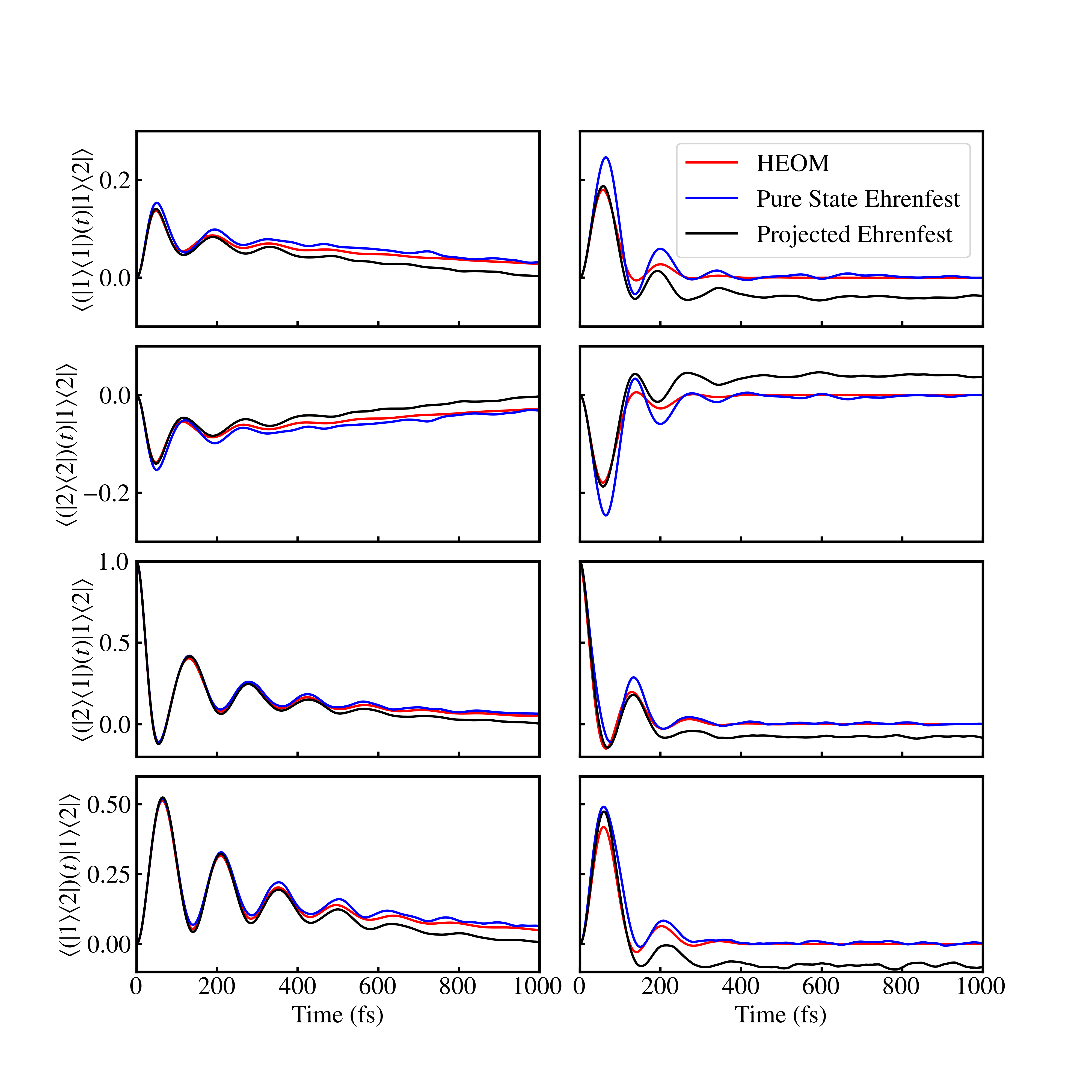}
    \vspace{-2mm}
    \caption{Single-time dynamics for an initial condition of $\ket{1}\bra{2}$ for both the slow (left) and fast (right) bath parameter regimes. The first two rows contain population dynamics (i.e. $\ket{1}\bra{1}$ and $\ket{2}\bra{2}$), while the bottom two rows show coherence dynamics ($\ket{2}\bra{1}$ and $\ket{1}\bra{2}$).}
    \vspace{-4mm}
    \label{fig:single-time-coh}
\end{figure}

The poor performance of the projected Ehrenfest method can be tracked down to discrepancies in the contributions of the response function. Figs.~\ref{fig:phi_2_real} and \ref{fig:phi_2_imag} illustrate this using plots of $\Phi_2$, with the rest of the contributions of the response function being shown in SI Figs.~\ref{fig:phi_1_slow}-~\ref{fig:phi_6_fast}. Here, we observe that, as expected, pure state Ehrenfest accurately reproduces the HEOM result while projected Ehrenfest fails to do so, and that the discrepancy is much more striking in the imaginary part of the response function. These discrepancies can be further traced to the single-time dynamics obtained from coherence initial conditions between excited states. Since the Frenkel exciton model is in the global ground state $\rho_0 = \ket{0}\bra{0}$ at $t=0$, such coherence states can only be obtained when the dipole operator is applied multiple times, as is the case when computing the third-order response. As such, the dynamics of coherence initial conditions between excited states (e.g. $\ket{1}\bra{2}$) are relevant to nonlinear spectra but not linear spectra. Figure~\ref{fig:single-time-coh} shows both the population and coherence dynamics for $\ket{1}\bra{2}$. In both the slow and fast bath parameter regimes, we see from the first two rows that the population dynamics obtained from pure state Ehrenfest more closely match the HEOM result, especially at longer times. In contrast, the projected Ehrenfest result matches the HEOM result until $\sim$200 fs, after which there are notable deviations. This poor performance of projected Ehrenfest is also observed to a lesser extent in the coherence dynamics of the $\ket{1}\bra{2}$ initial state (bottom two rows of Fig.~\ref{fig:single-time-coh}). We emphasize that the coherence dynamics presented here are distinct from those involved in the absorption spectra since \textit{they do not involve the ground state}. The worsening performance of projected Ehrenfest with time in reproducing the dynamics arising from $\ket{1}\bra{2}$ also explains why the respective 2DES spectrum degrades progressively as the waiting time $t_2$ increases (Fig.~\ref{fig:2D_spec_slow}). As such, the differences in the performance between projected and pure state Ehrenfest can be attributed to inaccuracies in the projected Ehrenfest dynamics when populations, and coherences to a lesser extent, arise from an initial condition of coherences between two excited states.

\begin{figure}[h!]
    \centering
    \includegraphics[width=0.4\textwidth]{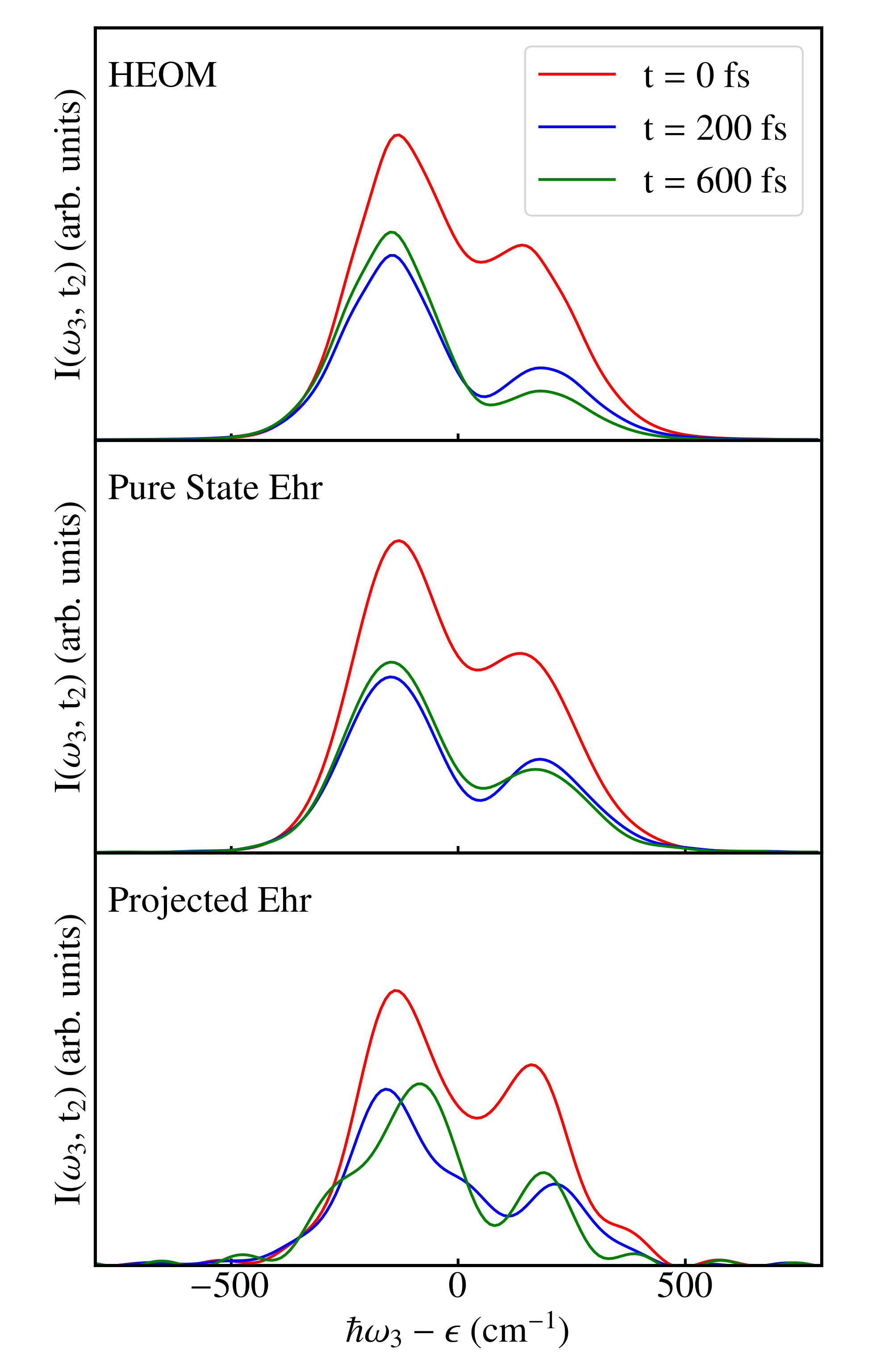}
    \caption{Pump-probe spectra for the slow bath parameter regime obtained from integrating the 2DES spectra over $\omega_1$.}
    \label{fig:pump_probe_slow}
\end{figure}

Pump-probe spectra, generated by integrating over $\omega_1$, are shown in Fig.~\ref{fig:pump_probe_slow}. The results here mirror those of the 2DES spectrum, where the pure state Ehrenfest method accurately reproduces the HEOM result, albeit with slightly higher peaks at 200 cm$^{-1}$ for $t_2 \geq$ 200 fs. In contrast, while the projected Ehrenfest method yields accurate results at $t_2 = 0$, the accuracy of the pump-probe spectra degrades at $t_2 \geq$ 200 fs, with the peak around 200 cm$^{-1}$ losing definition.

\begin{figure}[h!]
    \centering
    \includegraphics[width=0.5\textwidth]{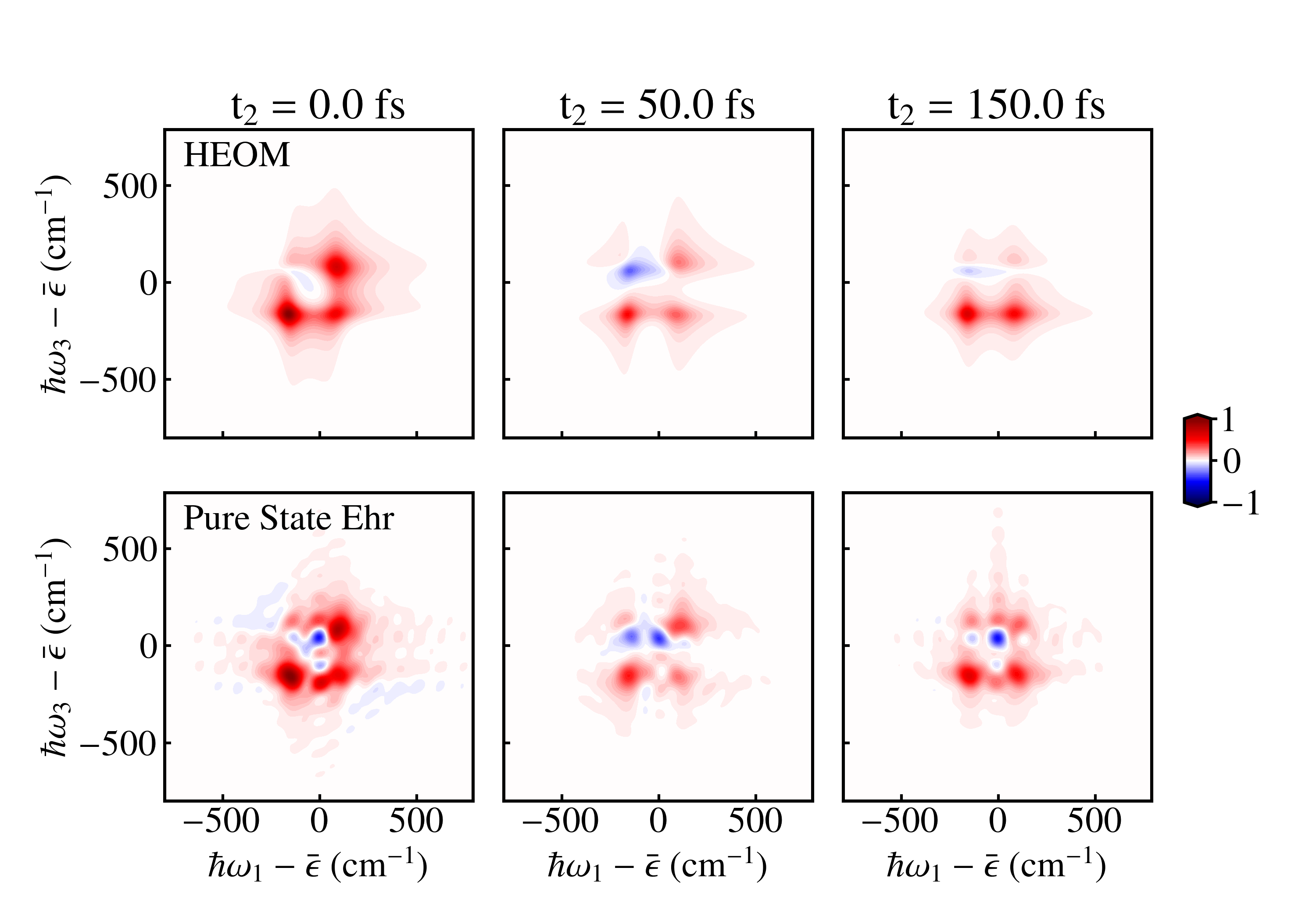}
    \vspace{-4mm}
    \caption{Exact two-dimensional spectra for the fast bath parameters computed using HEOM (top row), compared to the pure state Ehrenfest method (bottom row). All spectra are normalized such that the maximum amplitude equals 1.}
    \label{fig:fast_bath_2d}
\end{figure}

\begin{figure}[h!]
    \centering
    \vspace{-4mm}
    \includegraphics[width=0.4\textwidth]{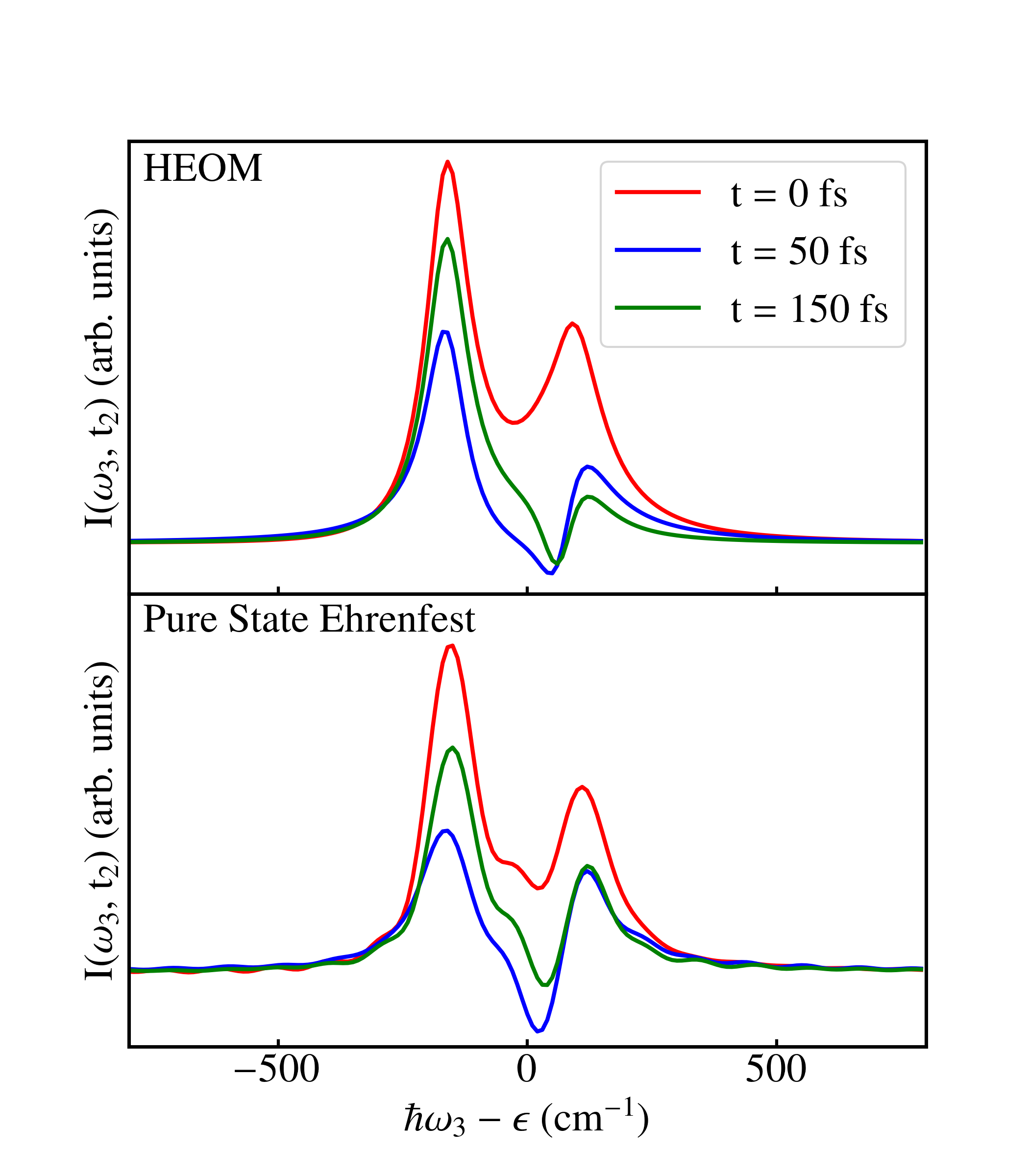}
    \vspace{-3mm}
    \caption{Pump-probe spectra for the fast bath parameter regime obtained from integrating the 2DES spectra over $\omega_1$.}
    \label{fig:pump_probe_fast}
\end{figure}

Next we examine how well pure state Ehrenfest performs in a more challenging fast bath regime. Fig.~\ref{fig:fast_bath_2d} shows 2DES spectra for the fast bath parameters computed via HEOM and pure state Ehrenfest. The range of values for $t_2$ (0 - 150 fs) is more limited here than in the slow bath regime because we observed virtually no change in the 2DES spectrum beyond $t_2=$200 fs (see SI Figure~\ref{fig:fast_bath_extended}). As with the slow bath parameter regime, the pure state Ehrenfest response functions were truncated before being Fourier transformed to produce the 2DES spectra (see SI Sec.~\ref{sec:third_order_truncation}). In this more challenging regime, pure state Ehrenfest does not perform as well as it did for the slow bath parameters. This is not surprising, as it aligns with the well-known deficiencies of the Ehrenfest method, which struggles to capture the correct dynamics in the limit of fast nuclei. Nonetheless, it is able to qualitatively capture most of the features present in the HEOM spectrum. This result is corroborated by the pump-probe spectra shown in Fig.~\ref{fig:pump_probe_fast}.

\section{Conclusion} \label{sec:conclude}
Here we have introduced the pure state Ehrenfest method to obtain linear and nonlinear spectra. We have shown that exploiting the linearity of the density matrix to express all initial conditions as sums of pure states is essential to obtaining accurate coherence dynamics. In particular, while our approach provides modest improvements over projected Ehrenfest in computing linear spectra, it is able to dramatically improve the accuracy with which nonlinear spectra (2DES and pump-probe) can be obtained, especially at longer waiting times. The physical origin of this improvement arises from the ability of our pure state Ehrenfest approach to much more accurately capture the dynamics when the initial condition is a coherence between excited states. These conditions do not occur when calculating the linear spectra (where all the initial density matrices are coherences with the ground state) but play a vital role in obtaining the higher-order response functions needed to capture nonlinear spectroscopy. We have shown that pure state Ehrenfest gives excellent agreement with the exact linear and 2DES spectra in the slow bath regime where the Ehrenfest method is expected to be accurate and have also demonstrated that it still provides good results in a faster bath regime and preserves the qualitative features observed in the exact spectra. These results suggest that the pure state Ehrenfest approach provides a tractable and accurate approach to calculate nonlinear spectra for multichromophoric systems in a wide range of chemical environments.

\section*{References}
\bibliography{references}

\section*{Acknowledgements}
This work was funded by the U.S. Department of Energy, Office of Science, Office of Basic Energy Sciences (DE-SC0020203 to T.E.M.). A.O.A. acknowledges support from the Edward Curtis Franklin Award and the Stanford Diversifying Academia, Recruiting Excellence Fellowship. A.M.C.~acknowledges the start-up funds from the University of Colorado, Boulder. 

\end{document}

\title{Supplementary Information for: An accurate and efficient Ehrenfest dynamics approach for calculating linear and nonlinear electronic spectra}

\author{Austin O. Atsango} 
\affiliation{Department of Chemistry, Stanford University, Stanford, California 94305, USA}

\author{Andr\'es Montoya-Castillo}
\email{Andres.MontoyaCastillo@colorado.edu}
\affiliation{Department of Chemistry, University of Colorado Boulder, Boulder, Colorado 80309, USA}

\author{Thomas E. Markland}
\email{tmarkland@stanford.edu}
\affiliation{Department of Chemistry, Stanford University, Stanford, California 94305, USA}

\date{\today}

\maketitle

\section{Third-Order Response Truncation} \label{sec:third_order_truncation}
For both the projected and pure state Ehrenfest methods, the third-order response functions, which are Fourier-transformed to produce the 2DES spectra, are truncated in $t_1$ and $t_3$ in order to exclude oscillations that occur in the time domain after the response function has decayed. This is because the oscillations degrade the quality of the resultant 2DES spectra. We applied the same cutoff to $t_1$ and $t_3$, i.e., $\textrm{max}(t_1)=\textrm{max}(t_3)=200$ fs for the slow bath parameter regime and $\textrm{max}(t_1)=\textrm{max}(t_3)=300$ fs for the fast bath parameter regime. Truncating $t_1$ and $t_3$ yielded 2DES spectra that were less noisy compared to those obtained from using damping functions that decay steadily from the cutoff.

\begin{figure}[h!]
    \centering
    \includegraphics[width=.7\textwidth]{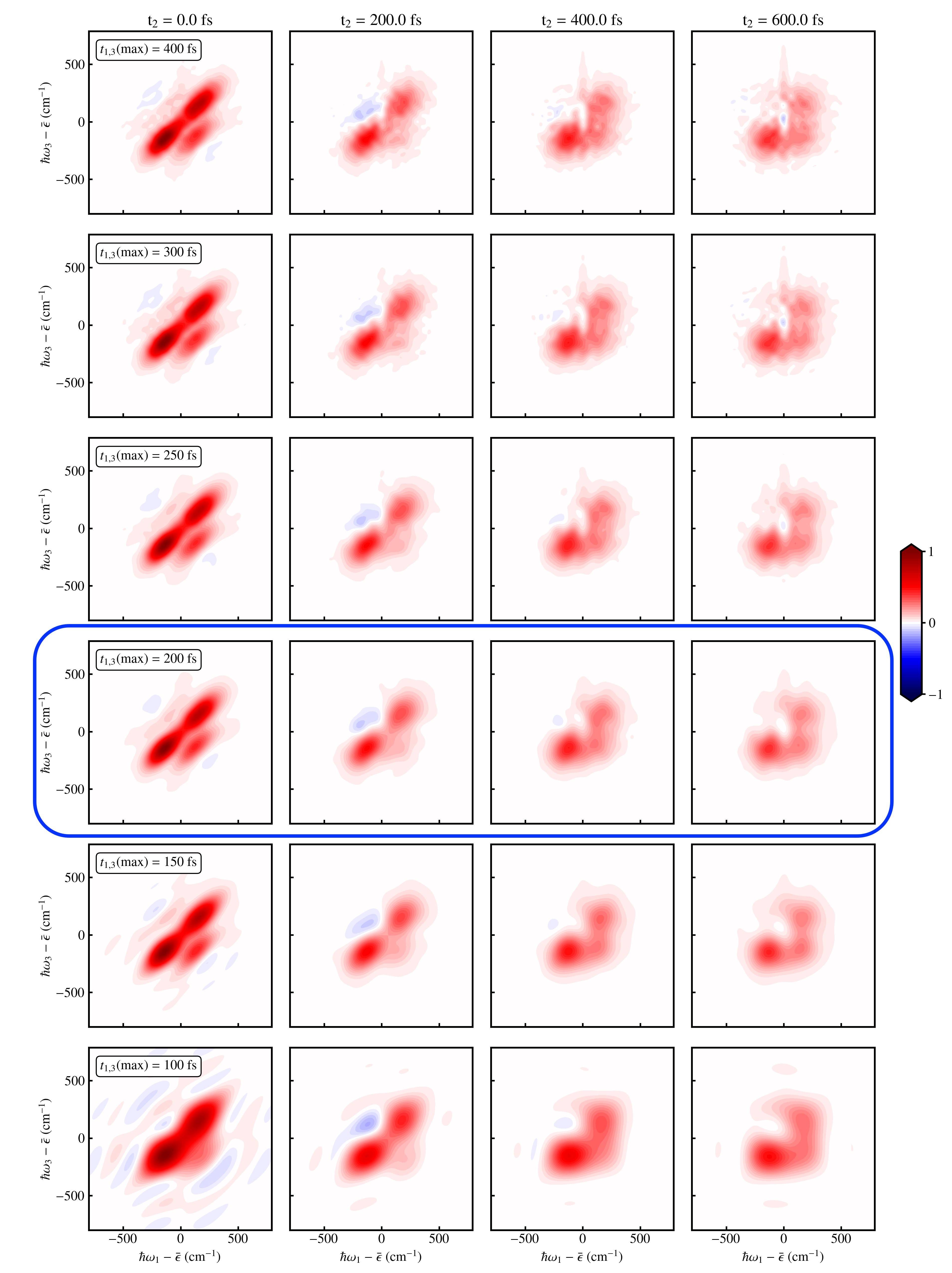}
    \caption{2DES spectra for the slow bath parameter regime at different final times for $t_1$ and $t_3$. Spectra for the chosen maximum $t_1=t_3 = 200$ fs are highlighted by the blue bounding box; these are presented in the main text.}
    \label{fig:spec_truncated}
\end{figure}

Figure~\ref{fig:spec_truncated} shows 2DES spectra obtained at different time cutoffs for the slow bath parameter regime. We observe that the spectra change progressively from generally noisy at the longer cutoff time (400fs) to smooth at 200 fs, which is the chosen time cutoff. Below 200 fs, the spectra start to become inaccurate because the third-order response function is now being cut off before it has decayed completely. Although the choice of $\textrm{max}(t_1)=\textrm{max}(t_3)$ for both the slow and fast bath parameter regimes was primarily motivated by the need to clean up the spectra, one can pick these values based on the single time-coherence dynamics shown in Fig.~\ref{fig:coh_dynamics} in the main text. There, we see that setting the maximum $t_1$ as 200 fs and 300 fs for the slow and fast bath parameters respectively is justified because all spectral features fall below the chosen time limit in both cases, especially for the more accurate pure state Ehrenfest and HEOM methods.

\clearpage
\section{Third-Order Response Function Components for Slow Bath Parameters}

Here, we show the rephasing and non-rephasing components of the third-order response function as detailed in Equations~\ref{eq:3o_response}-~\ref{eq:response_cont} for the slow bath parameter regime. The plots include the ground state bleaching, stimulated emission, and excited state absorption contributions.

\begin{figure}[h!]
    \centering
    \includegraphics[width=.8\textwidth]{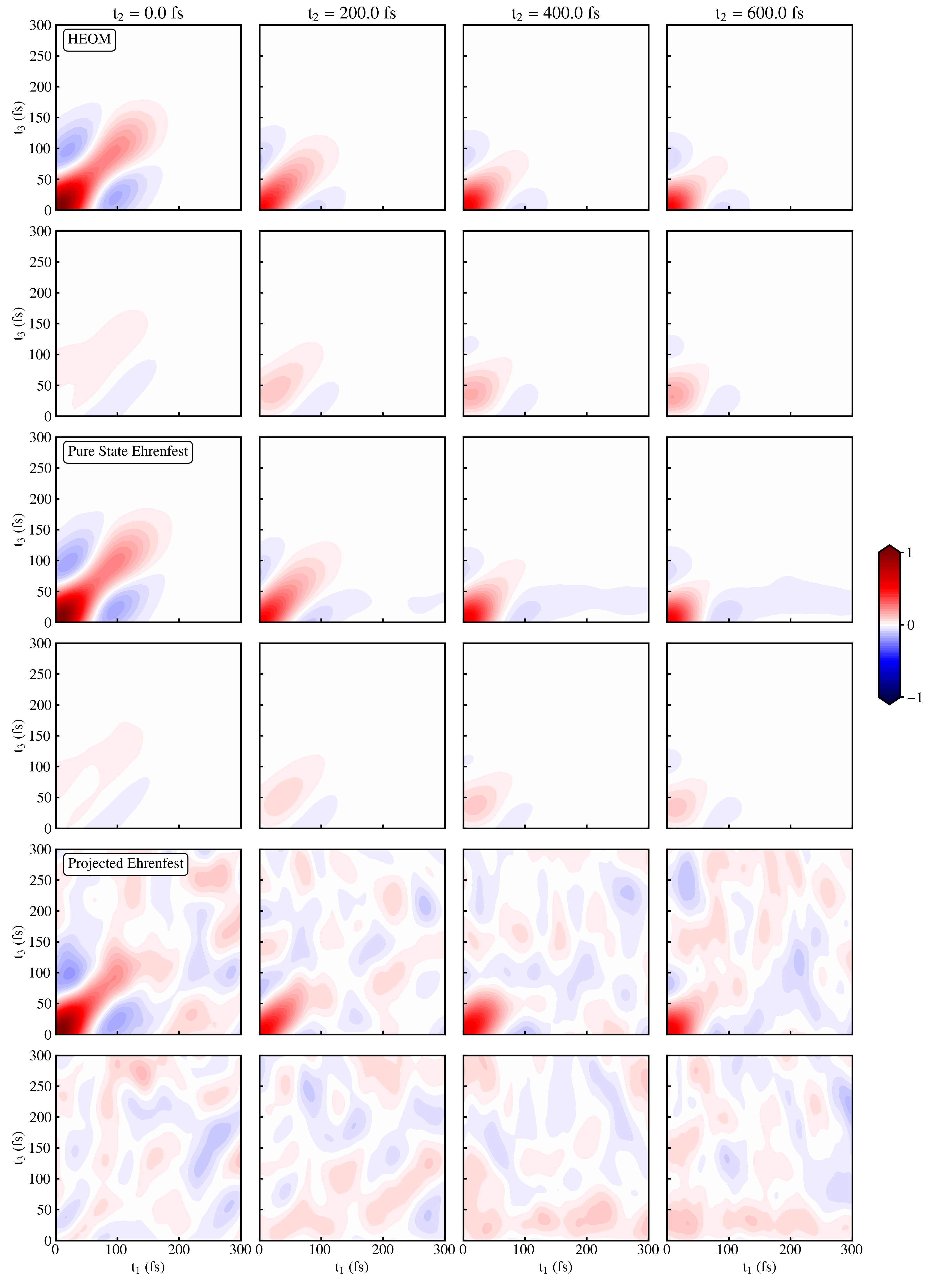}
    \caption{$\Phi_1$: Rephasing, stimulated emission}
    \label{fig:phi_1_slow}
\end{figure}

\begin{figure}[h!]
    \centering
    \includegraphics[width=.8\textwidth]{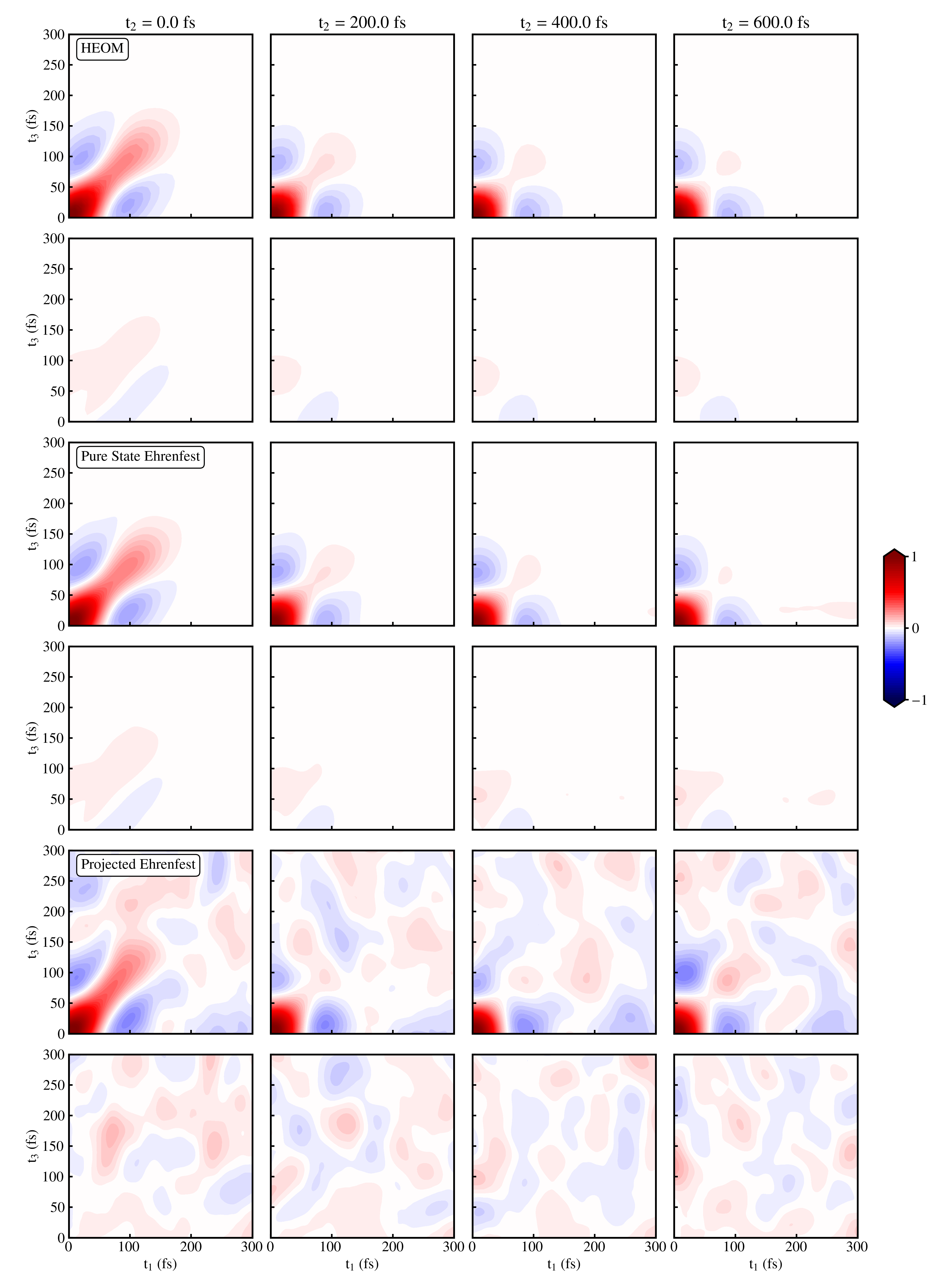}
    \caption{$\Phi_2$: Rephasing, ground state bleaching}
    \label{fig:phi_2_slow}
\end{figure}

\begin{figure}[h!]
    \centering
    \includegraphics[width=.8\textwidth]{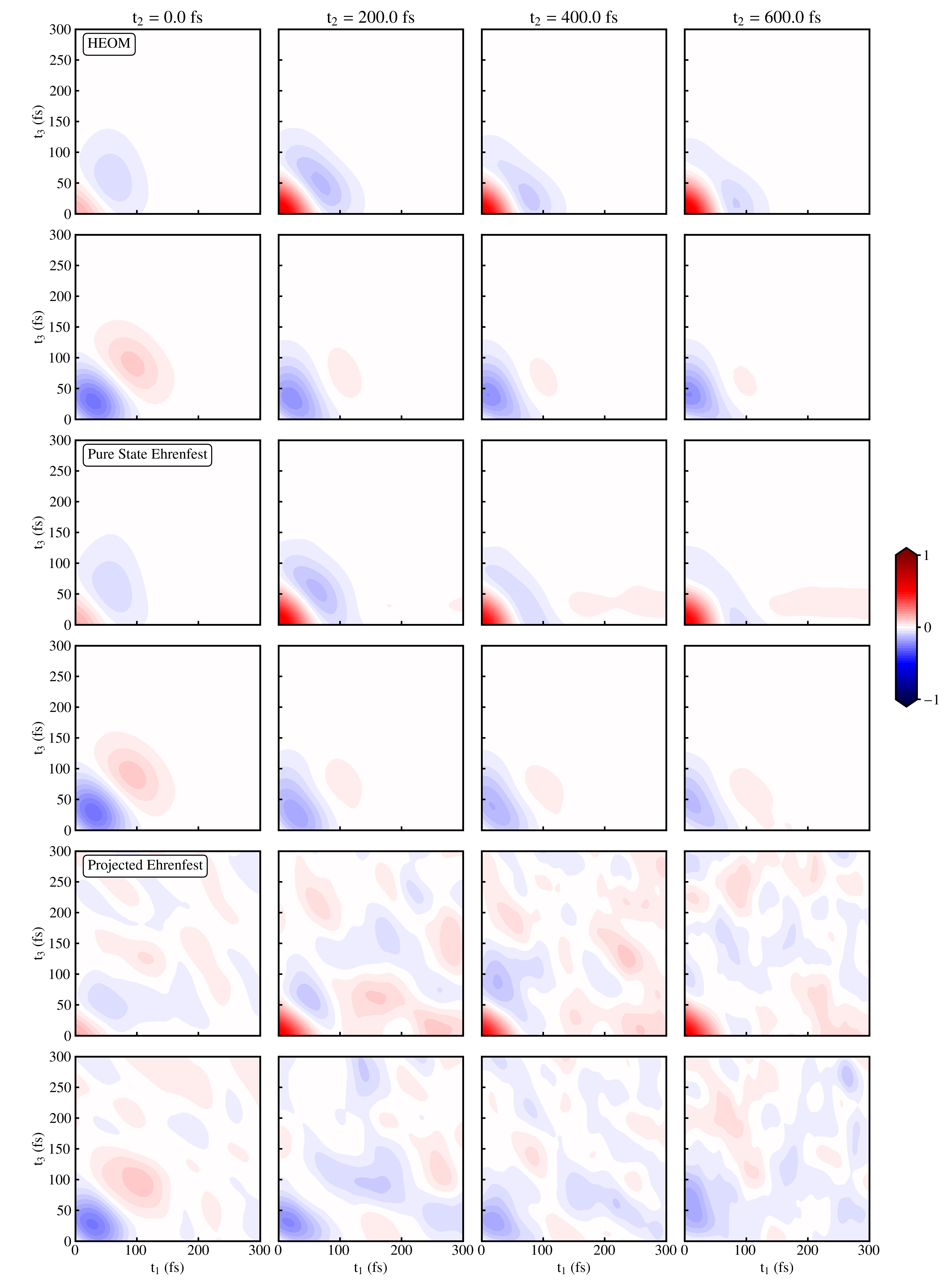}
    \caption{$\Phi_3$: Rephasing, excited state absorption}
    \label{fig:phi_3_slow}
\end{figure}

\begin{figure}[h!]
    \centering
    \includegraphics[width=.8\textwidth]{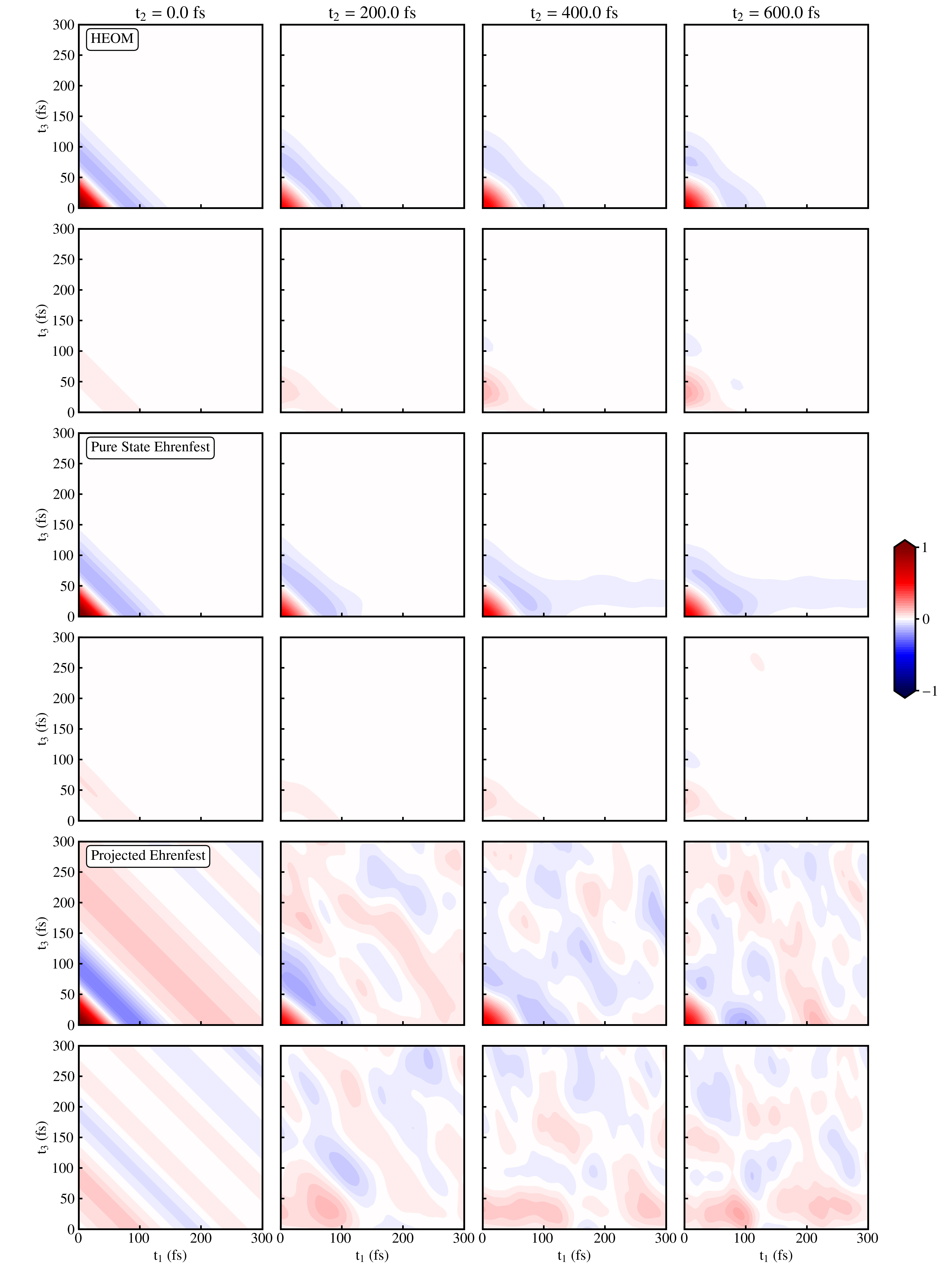}
    \caption{$\Phi_4$: Non-rephasing, stimulated emission}
    \label{fig:phi_4_slow}
\end{figure}

\begin{figure}[h!]
    \centering
    \includegraphics[width=.8\textwidth]{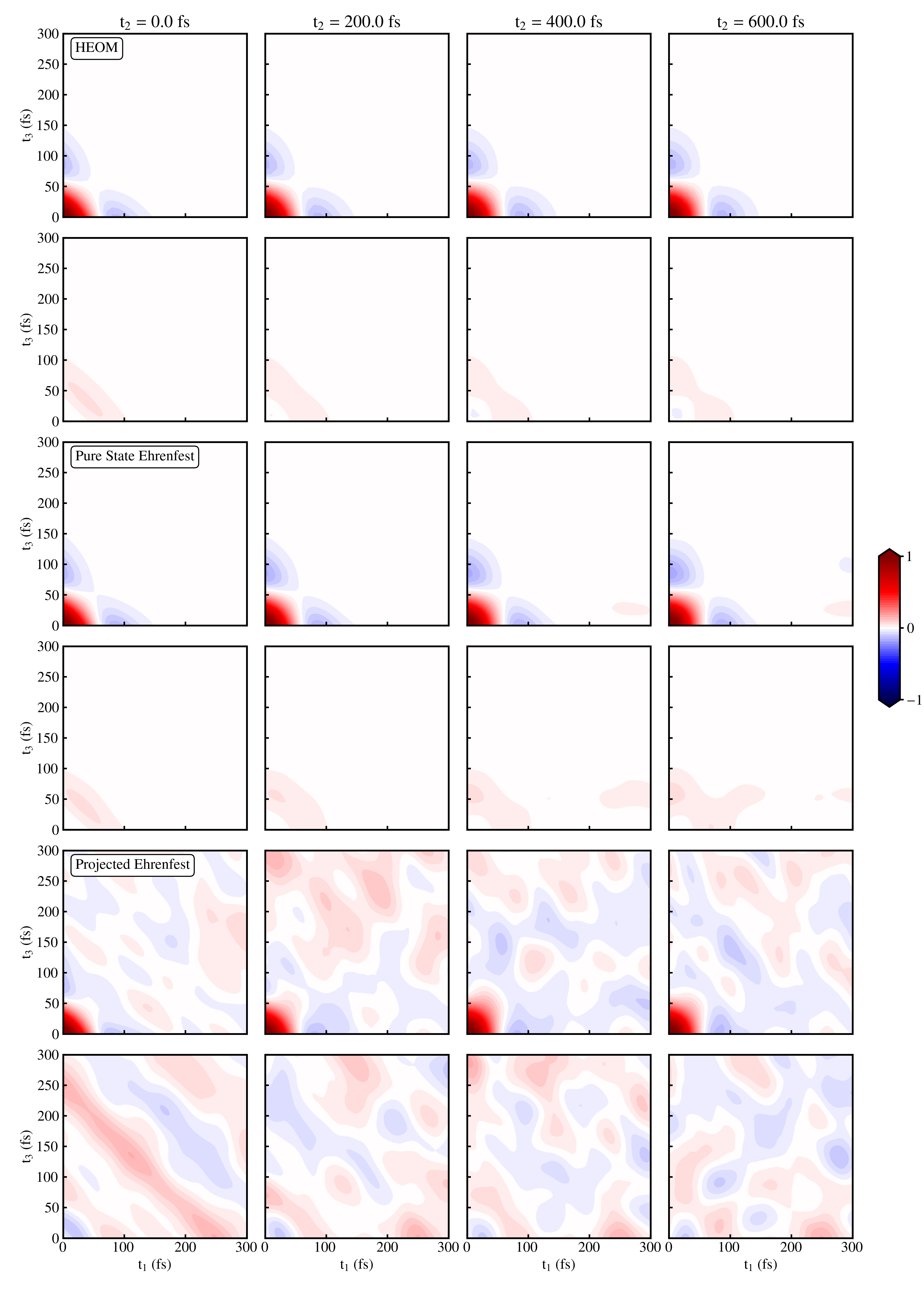}
    \caption{$\Phi_5$: Non-rephasing, ground state bleaching}
    \label{fig:phi_5_slow}
\end{figure}

\begin{figure}[h!]
    \centering
    \includegraphics[width=.8\textwidth]{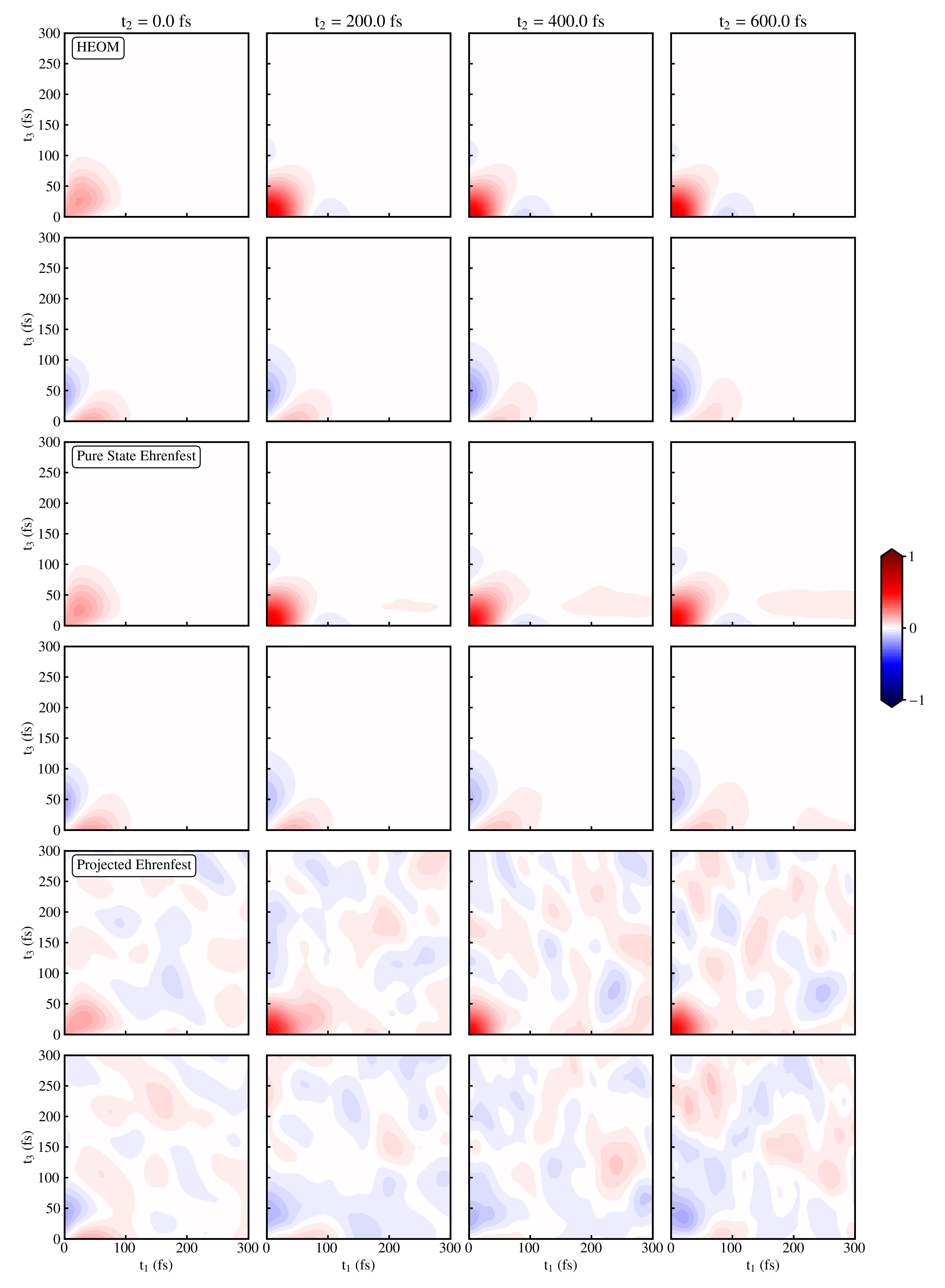}
    \caption{$\Phi_6$: Non-rephasing, excited state absorption}
    \label{fig:phi_6_slow}
\end{figure}

\clearpage

\section{Third-Order Response Function Components for Fast Bath Parameters}
Here, we show the rephasing and non-rephasing components of the third-order response function for the fast bath parameter regime. The plots include the ground state bleaching, stimulated emission, and excited state absorption contributions.

\begin{figure}[h!]
    \centering
    \includegraphics[width=.9\textwidth]{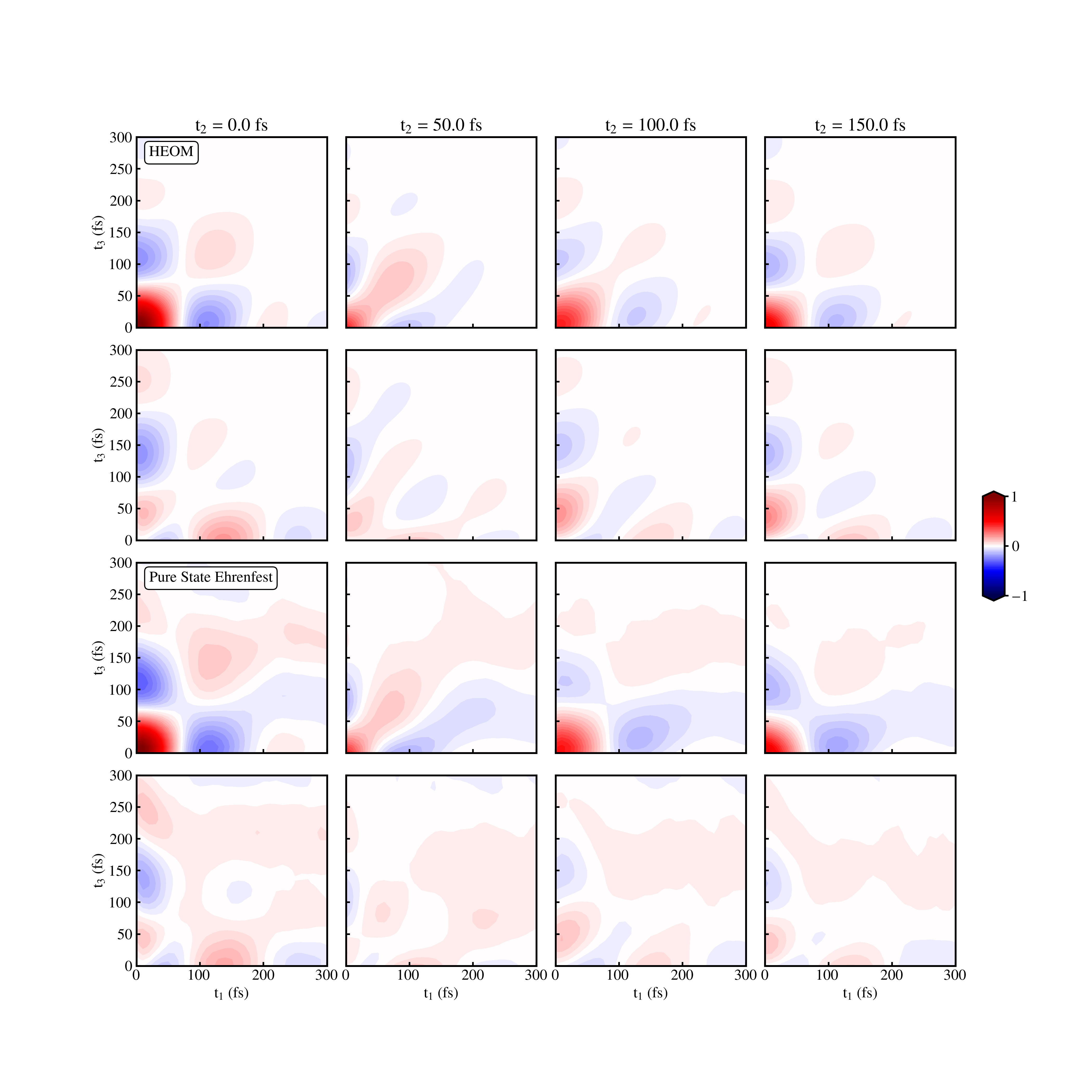}
    \caption{$\Phi_1$: Rephasing, stimulated emission}
    \label{fig:phi_1_fast}
\end{figure}

\begin{figure}[h!]
    \centering
    \includegraphics[width=.9\textwidth]{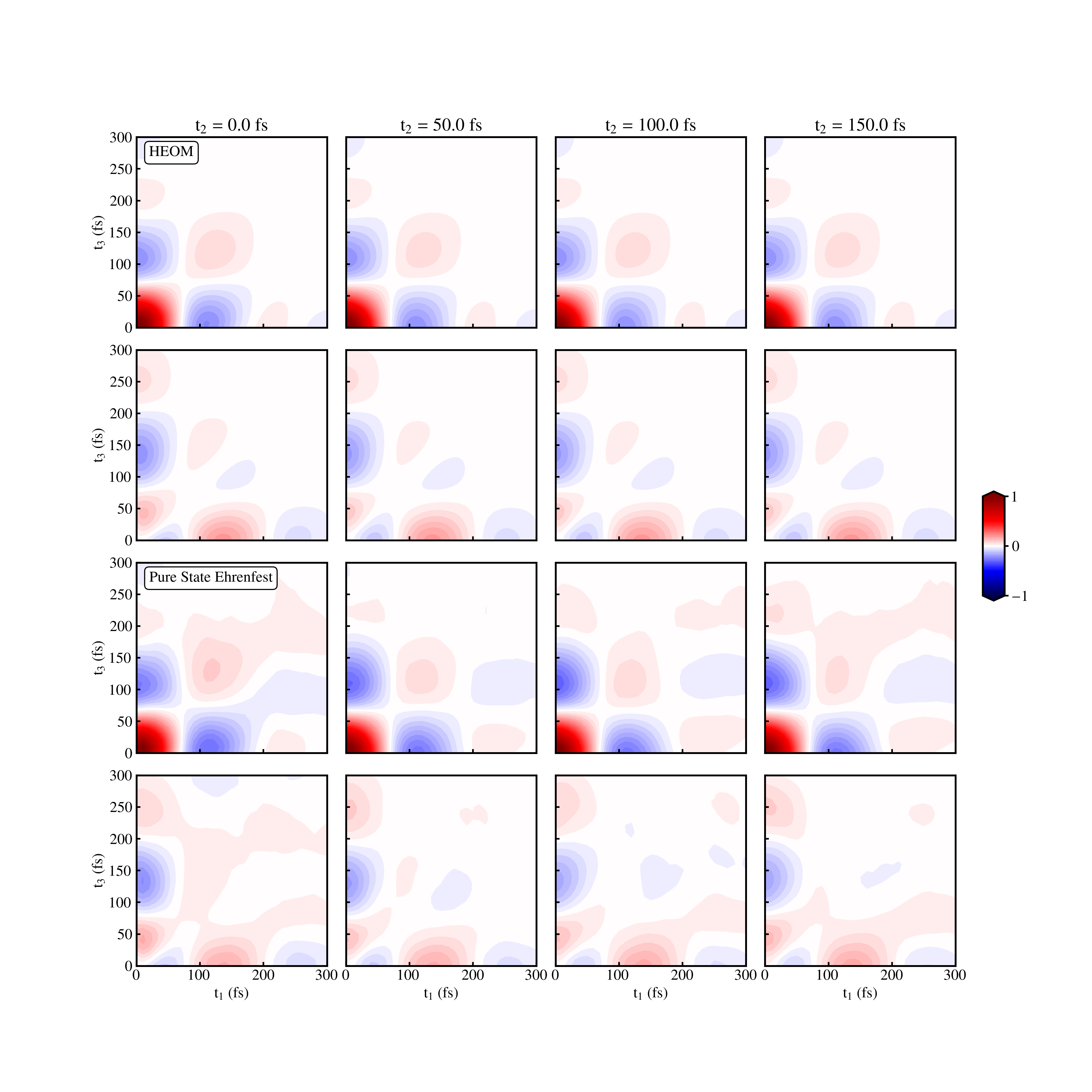}
    \caption{$\Phi_2$: Rephasing, ground state bleaching}
    \label{fig:phi_2_fast}
\end{figure}

\begin{figure}[h!]
    \centering
    \includegraphics[width=.9\textwidth]{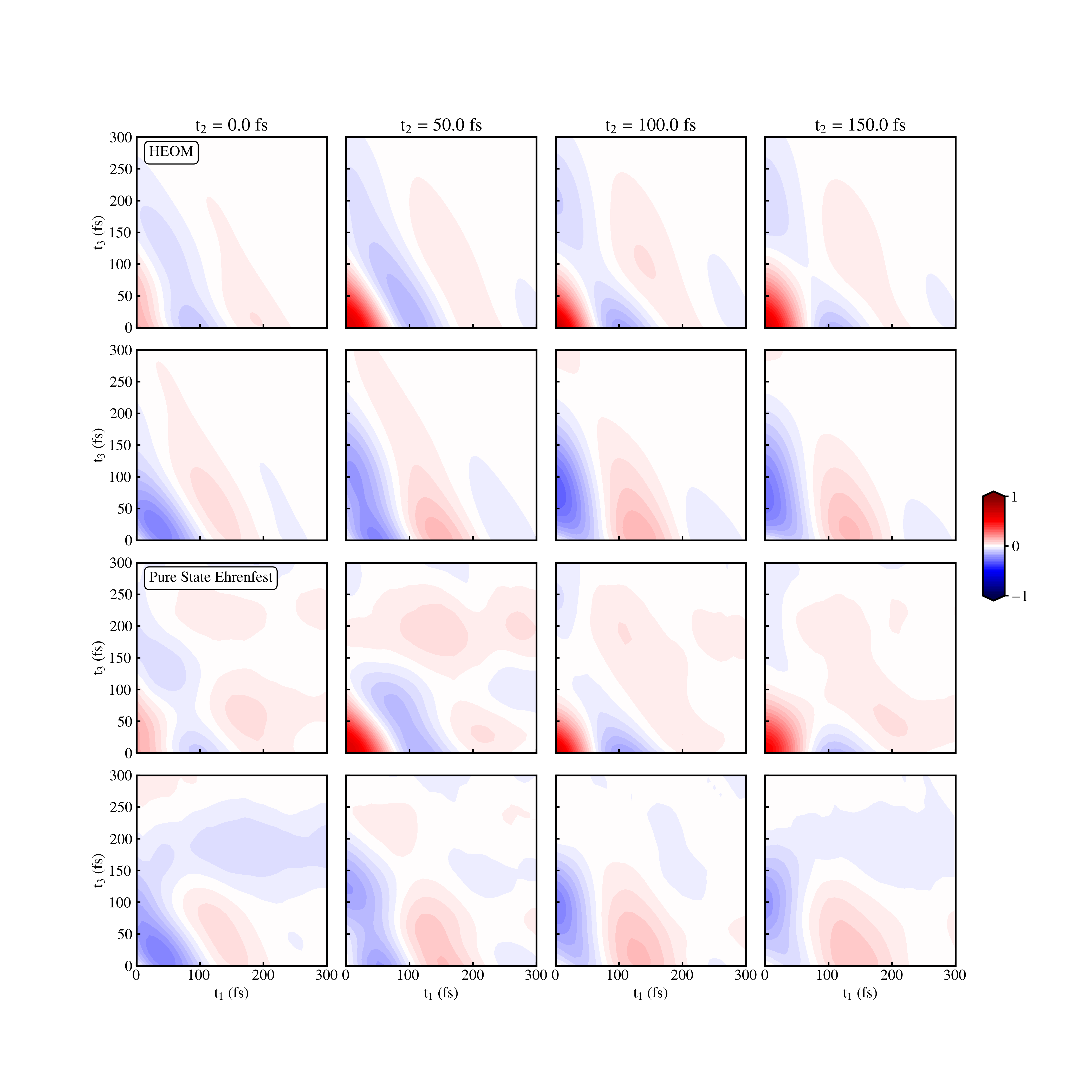}
    \caption{$\Phi_3$: Rephasing, excited state absorption}
    \label{fig:phi_3_fast}
\end{figure}

\begin{figure}[h!]
    \centering
    \includegraphics[width=.9\textwidth]{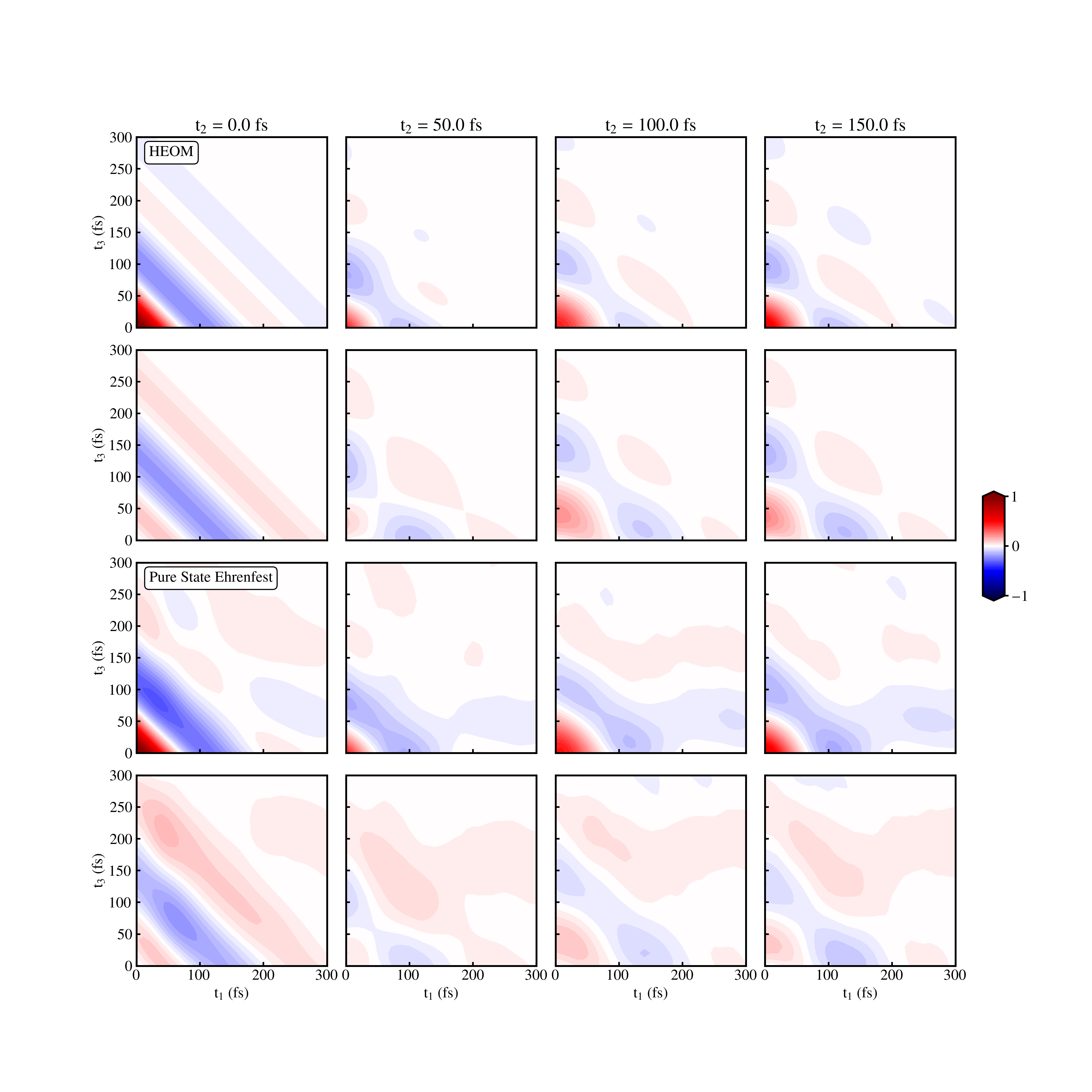}
    \caption{$\Phi_4$: Nonrephasing, stimulated emission}
    \label{fig:phi_4_fast}
\end{figure}

\begin{figure}[h!]
    \centering
    \includegraphics[width=.9\textwidth]{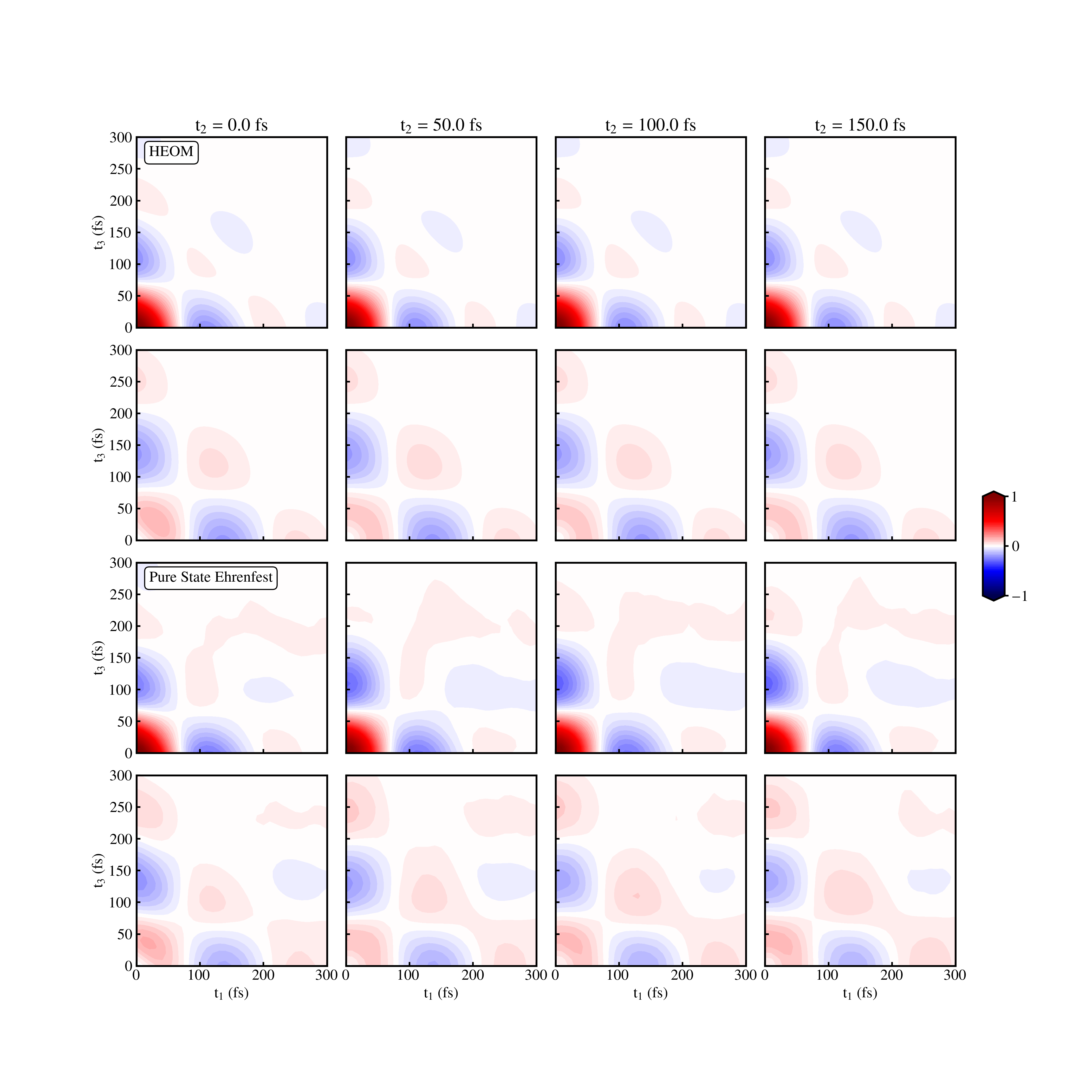}
    \caption{$\Phi_5$: Nonrephasing, ground state bleaching}
    \label{fig:phi_5_fast}
\end{figure}

\begin{figure}[h!]
    \centering
    \includegraphics[width=.9\textwidth]{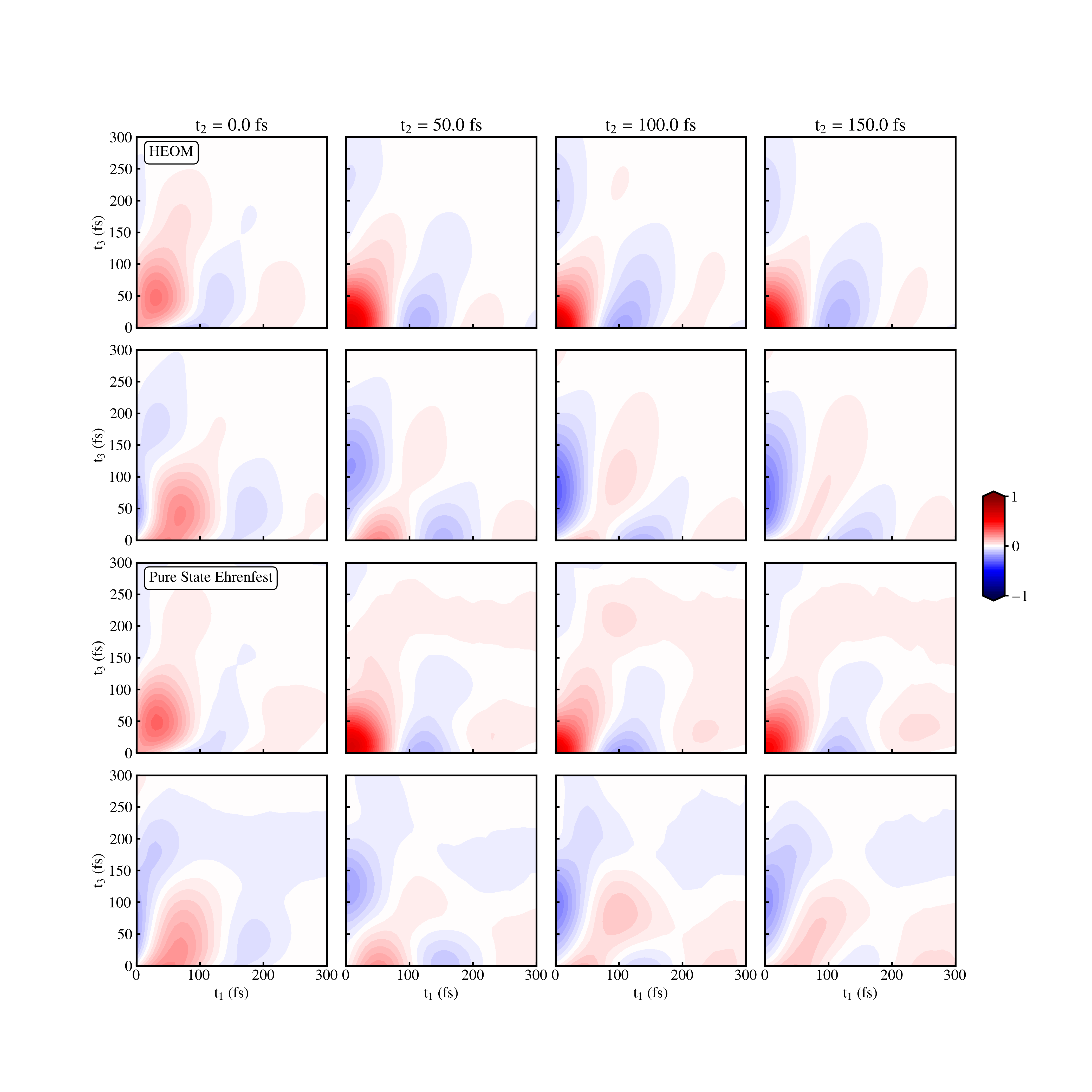}
    \caption{$\Phi_6$: Nonrephasing, excited state absorption}
    \label{fig:phi_6_fast}
\end{figure}

\clearpage

\section{Two-Dimensional Electronic Spectra for the Fast Bath Regime at Extended Waiting Times}

2DES spectra for the fast bath parameter regime in the main text were presented for waiting times $t_2$ = 0 - 150 fs. Here, we show the 2DES spectra at $t_2 = \{ 200 fs, 400 fs, 600fs\}$, which demonstrate that there is little change during this period, thus justifying the choice to focus on a more limited range of $t_2$, i.e. 0 -150 fs as shown in the main text.
\begin{figure}[h!]
    \centering
    \includegraphics[width=.9\textwidth]{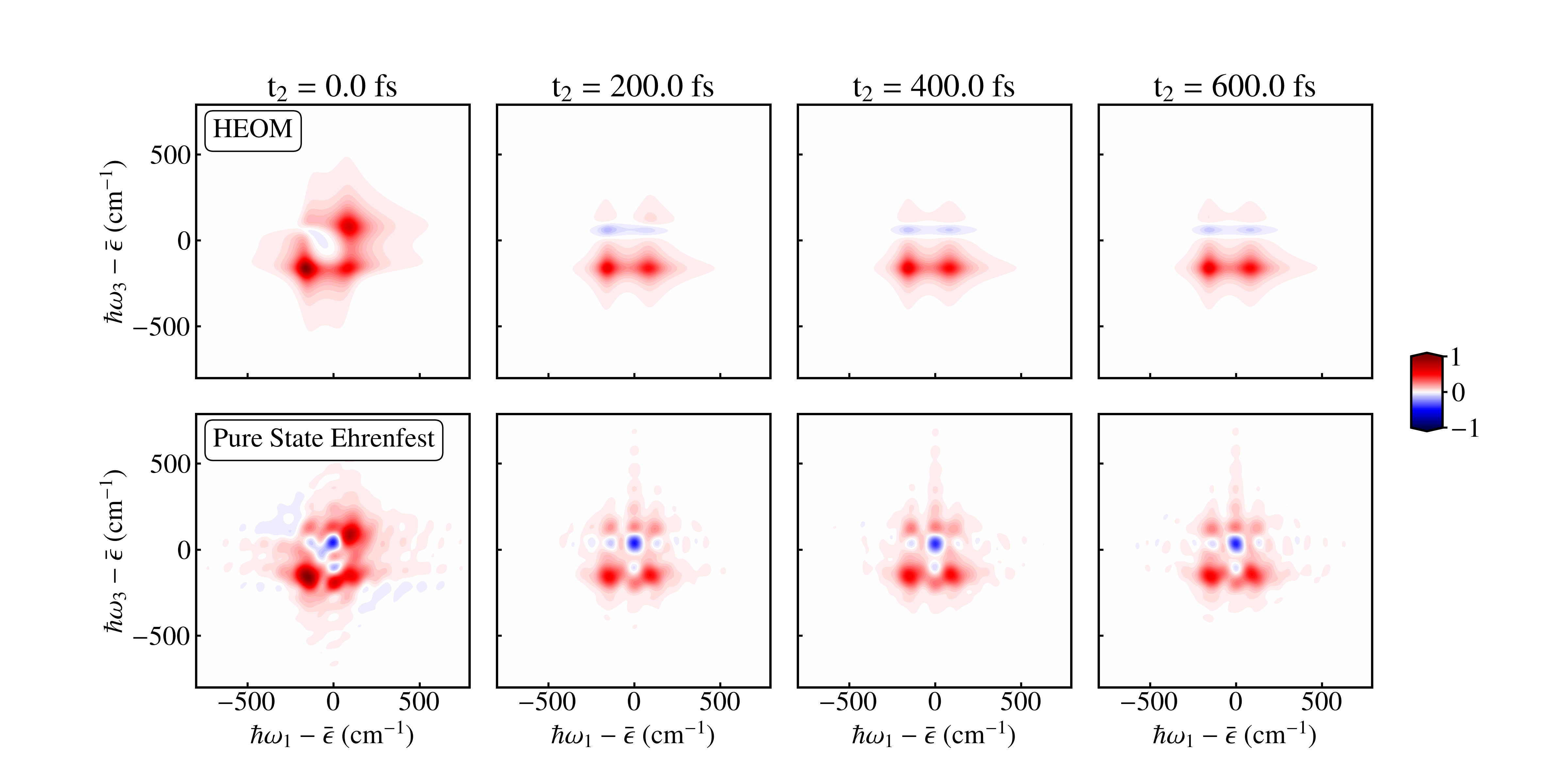}
    \caption{Exact two-dimensional spectra for the fast bath parameters computed using HEOM (top row), compared to the pure state Ehrenfest method (bottom row). As mentioned in the main text, there is little change in the spectra after $t_2 = 200$ fs.  All spectra are normalized such that the maximum amplitude equals 1.}
    \label{fig:fast_bath_extended}
\end{figure}